\documentclass[a4paper,11pt]{article}
\pdfoutput=1 % if your are submitting a pdflatex (i.e. if you have
\usepackage{jheppub} % for details on the use of the package, please
\usepackage[T1]{fontenc} % if needed
\usepackage{makecell}
\usepackage{hyperref}
\hypersetup{colorlinks=true
  ,urlcolor=blue
  ,anchorcolor=blue
  ,citecolor=blue
  ,filecolor=blue
  ,linkcolor=blue
  ,menucolor=blue
  ,pagecolor=blue
  ,linktocpage=true
  ,pdfproducer=medialab
  ,pdfa=true}

\title{ \bf 6d superconformal Cardy formulas}

\author[]{June Nahmgoong}

\affiliation[]{Department of Physics and Astronomy \& Center for Theoretical Physics\\ Seoul National University, Seoul 08826, Korea}

\emailAdd{junenahmgoong@gmail.com}

\abstract{We study the superconformal index of 6d SCFTs from their 't Hooft anomalies. In the Cardy limit where the angular momenta on \(S^5\) are large, we show that the leading free energy, as well as a few subleading corrections, can be computed from the 6d anomaly polynomials. Our large \(N\) free energy accounts for the entropy of supersymmetric black holes in dual \(AdS_7\). }

\preprint{SNUTP19-003}

\begin{document} 
\maketitle
\flushbottom

\section{Introduction and summary}
The superconformal index of a \(d\)-dimensional SCFT counts BPS states on \(S^{d-1}\) \cite{Kinney:2005ej,Romelsberger:2005eg,Bhattacharya:2008zy}. For a 6d SCFT, its superconformal index is defined as
\begin{align}
    \mathcal{I}=\text{Tr}\Bigr[(-1)^F \cdot e^{-\Delta_R Q_R} \cdot e^{-\omega_1 J_1-\omega_2 J_2-\omega_3 J_3} \cdot \prod_i e^{-m_i F_i} \Bigr]
    \label{eq56_index}
\end{align}
where \(J_{1,2,3}\) are the angular momenta on \(S^5\), \(Q_R\) is the Cartan charge of \(SU(2)\) R-symmetry of (1,0) superconformal algebra, and \(F_i\)'s are the charges of the other global symmetries. Chemical potentials \(\Delta_R\) and \(\omega_{1,2,3}\) are constrained by \(\Delta_R-\omega_1-\omega_2-\omega_3=0\), and only BPS states which saturate the BPS bound \(E\geq 4Q_R+J_1+J_2+J_3\) contribute to the index. The 6d superconformal index was studied in \cite{Kim:2012ava,Kallen:2012zn,Kim:2012tr,Lockhart:2012vp,Kim:2012qf,Minahan:2013jwa,Kim:2013nva} and see \cite{Kim:2016usy} for a detailed review. 
\par 
6d SCFTs on \(N\) M5-branes are expected to have \(N^3\) degrees of freedom \cite{Klebanov:1996un}. \(N^3\) dependence of 6d SCFTs are observed in the anomaly polynomials \cite{Freed:1998tg,Ohmori:2014kda,Intriligator:2014eaa}, the Casimir energy \cite{Kim:2012ava,Kim:2012qf,Minahan:2013jwa,Kim:2013nva,Bobev:2015kza}, and \(\mathbb{R}^4\times T^2\) partition function \cite{Kim:2011mv,Kim:2017zyo}. However, there has been no direct evidence for \(N^3\) growth of states from the superconformal index.
\par 
As a related matter, it had been believed that the superconformal index could not capture the entropy of BPS black holes in the dual AdS space due to the severe boson/fermion cancellation. However, inspired by the extremization principle found in \cite{Hosseini:2017mds,Hosseini:2018dob,Choi:2018fdc}, it has been recently observed that the superconformal index can actually reproduce the entropy of supersymmetric black holes in \(AdS_4\) \cite{Choi:2019zpz, Nian:2019pxj, Choi:2019dfu}, \(AdS_5\) \cite{Cabo-Bizet:2018ehj,Choi:2018hmj,Benini:2018ywd}, \(AdS_6\) \cite{Choi:2019miv}, and \(AdS_7\) \cite{Choi:2018hmj}. The primary breakthrough was allowing the chemical potentials to be complex to obstruct \((-1)^F\) cancellation \cite{Choi:2018vbz}. The complexified solution of the chemical potentials was recently obtained also in the gravity dual \cite{Cassani:2019mms,Kantor:2019lfo}. 
\par 
The Cardy formula of 2d CFT \cite{Cardy:1986ie} plays a crucial role in accounting the entropy of black holes from field theories \cite{Strominger:1996sh}. In \(d>2\), the Cardy formula refers to the behavior of the superconformal index in the `generalized Cardy limit' where \(|\omega_i|\ll 1\). It corresponds to the limit where the angular momenta are large, and therefore it is in analogy with the limit considered in the original Cardy formula of 2d CFT. In \cite{DiPietro:2014bca}, the Cardy formula of 4d SCFTs was studied with the strictly real chemical potentials. It was further studied in \cite{Choi:2018hmj,Honda:2019cio,ArabiArdehali:2019tdm,Kim:2019yrz,Cabo-Bizet:2019osg} at the complex chemical potentials.
It turns out that the free energy of the index captures \(N^2\) growth, which accounts for the entropy of \(AdS_5\) black holes. 
\par 
The 6d Cardy formula was obtained in \cite{DiPietro:2014bca, Chang:2019uag} in the setting where \(N^3\) growth is invisible. With the complex chemical potentials, a similar study of 6d SCFTs was initiated by \cite{Choi:2018hmj} for (2,0) \(A_N\) theories in the large \(N\) limit to reproduce the entropy of \(AdS_7\) black holes. In this paper, we make comprehensive studies of various 6d SCFTs. In particular, we will complete the argument given in \cite{Choi:2018hmj}, and obtain Cardy formulas of various 6d SCFTs.
\par 
In this paper, we study the modified version of the superconformal index which is defined as follows,
\begin{align}
I=\text{Tr}\Bigr[e^{-\Delta_R  Q_R}\cdot e^{-\omega_1 J_1-\omega_2 J_2-\omega_3 J_3} \cdot \prod_i e^{-m_i F_i} \Bigr]
\label{eq65_modind}
\end{align}
where chemical potentials are constrained by \(\Delta_R-\omega_1-\omega_2-\omega_3=2\pi i\). The modified index counts the same BPS states with the original index but on the different chemical potential basis. Note that \((-1)^F=e^{2\pi i J_3}\) since \(J_I\)'s are normalized to be integers for bosonic states and half-integers for fermionic states. Then, we can replace \((-1)^F\) in (\ref{eq56_index}) by \(e^{2\pi i J_3}\) and then absorb it into the redefinition of \(\omega_3\) which shifts the chemical potential constraint. The modified index turns out to be more `refined' in the sense that it can capture more entropy than the original index in the Cardy limit. Similar modification of the index in 4d was studied in \cite{Kim:2019yrz}.
\par 
In order to compute the index (\ref{eq65_modind}), we consider the following thermal partition function on \(S^5\times S^1\),
\begin{align}
Z=\text{Tr}\Bigr[e^{-\beta E } \cdot e^{-\Delta_R Q_R}\cdot e^{-\omega_1 J_1-\omega_2 J_2-\omega_3 J_3} \cdot \prod_i e^{-m_i F_i} \Bigr]
\label{eq69_part}
\end{align}
where the periodicity of the temporal \(S^1\) is given by \(\beta\). The partition function (\ref{eq69_part}) becomes the index (\ref{eq65_modind}) if one takes \(\beta\to 0\) limit and imposes \(\Delta_R-\omega_1-\omega_2-\omega_3=2\pi i\) constraint \cite{Choi:2018hmj}. The partition function can be computed from the effective action of background fields, which depend on the chemical potentials. In \(\beta\to 0\) limit, the infinitely many terms in the effective action are arranged into a derivative expansion on \(S^5\). Among them, Chern-Simons terms can be entirely determined by the anomaly polynomials \cite{Jensen:2012kj,Jensen:2013kka,Jensen:2013rga}. We will show that in the Cardy limit \(|\omega_i|\ll 1\), the leading free energy of the index, as well as a few subleading corrections, are determined from the Chern-Simons action only.
\par 
For a general 6d SCFT, we find that the free energy (\(\log I\)) of the modified index (\ref{eq65_modind}) with \(m_i=0\) is given as follows,
\begin{align}
\log I=&-\Bigr({ \mathfrak{a} \over 384}\Delta_R^4 +{ \mathfrak{b} \pi^2\over 24} \Delta_R^2+{2 \mathfrak{c} \pi^4\over 3} \Bigr){1\over \omega_1 \omega_2 \omega_3}
+\Bigr({ \mathfrak{b} \over 96}\Delta_R^2 +{  (2\mathfrak{c}+\mathfrak{d}) \pi^2\over 6}\Bigr){\omega_1^2 +\omega_2^2+\omega_3^2\over \omega_1 \omega_2 \omega_3}
\nonumber \\
&+\mathcal{O}(\log \omega).
\label{eq89_general}
\end{align}
The series expansion parameter is given by \(\omega\) which is the common scaling factor of \(\omega_I\)'s in the Cardy limit, i.e., \(|\omega_I|\sim\omega \ll 1\). Here \(\mathfrak{a,b,c}\) and \(\mathfrak{d}\) are the 't Hooft anomaly coefficients defined as follows,
\begin{align}
    P_8={1\over 4!}\Bigr(\mathfrak{a}\cdot c_2(R)^2+\mathfrak{b}\cdot c_2(R) p_1(T)+\mathfrak{c} \cdot p_1(T)^2+\mathfrak{d} \cdot p_2(T)  \Bigr)
    \label{eq95_anom}
\end{align}
where \(P_8\) is the anomaly 8-form of the 6d SCFT. Here, \(c_2(R)\) is the second Chern class of \(SU(2)\) R-symmetry and \(p_{1,2}(T)\) are the Pontryagin classes of the tangent bundle. Our Cardy formula (\ref{eq89_general}) takes the form of the `Cardy series' in the sense that it determines the free energy up to the subleading \(\mathcal{O}(\omega^{-1})\) order. If we perform Legendre transformation of the free energy, we obtain the asymptotic entropy in the Cardy limit. We found that the real part of the entropy is non-negative as long as the 't Hooft anomaly coefficients satisfy the following bound,
\begin{align}
\mathfrak{a}-4\mathfrak{b}+16\mathfrak{c} \geq 0
\label{eq99_ineq}
\end{align}
which is satisfied for all examples analyzed in this paper. The same form of the inequality (\ref{eq99_ineq}) was also found from the Renyi entropy of 6d SCFTs \cite{Yankielowicz:2017xkf}.
\par 
For 6d (2,0) SCFT of ADE type, our Cardy formula determines the free energy as follows,
\begin{align}
&\log I=-{h_G^\vee d_G+r_G\over 384}{(\Delta_R^2-\Delta_L^2)^2\over \omega_1 \omega_2 \omega_3}
\nonumber \\
&-r_G\Bigr[{(\Delta_R^2 +4\pi^2)( \Delta_L^2+4\pi^2)\over 192\omega_1 \omega_2 \omega_3}-{\Delta_R^2+\Delta_L^2-8\pi^2\over 192}{\omega_1^2+\omega_2^2+\omega_3^2\over \omega_1 \omega_2 \omega_3} \Bigr]+\mathcal{O}(\log \omega)
\label{eq78_cardy}
\end{align}
where \(\Delta_{L,R}\) are chemical potentials for \(SU(2)_L\times SU(2)_R\subset SO(5)\) R-symmetry, and chemical potentials are constrained by \(\Delta_R-\omega_1-\omega_2-\omega_3=2\pi i\). Here, \(\omega\) denotes a common scale for \(\omega_{1,2,3}\) which is a small parameter. The group theoretic constants \(h_G^\vee, d_G\) and \(r_G\) are listed in table \ref{tab_group_const}. In the large \(N\) limit of \(A_N\) type SCFT, the free energy (\ref{eq78_cardy}) shows explicit \(N^3\) growth which accounts for the entropy of BPS black holes in \(AdS_7\times S^4\) \cite{Hosseini:2018dob,Choi:2018hmj}. We make the similar analysis for the \(D_N\) type theory in the large \(N\) limit and found the correspondence with the entropy of BPS black holes in \(AdS_7\times S^4/\mathbb{Z}_2\).
\par 
The supersymmetric Casimir energy takes the same form with the equivariant integral of the anomaly polynomial \cite{Bobev:2015kza}. For the Cardy free energy, we found that it can be obtained from the equivariant integral of the `thermal anomaly polynomial' as follows,
\begin{align}
\log I = - \int P_8^T+\mathcal{O}(\log \omega).
\end{align}
Here, \(P_8^T\) is the thermal anomaly polynomial of the 6d SCFT defined from the following replacement rule \cite{Jensen:2013rga},
\begin{align}
P_8^T=P_8\Bigr(p_k(T)\to p_k(T)-{\mathbf{F}_T^2\over 4\pi^2}p_{k-1}(T) \Bigr)
\end{align}
where \(\mathbf{F}_T\) is a field strength of the fictitious gauge field. The detailed explanation will be given in section \ref{section2}.
\par 
The rest of the paper is organized as follows. In section \ref{section2}, we review the thermal derivative expansion of a thermal partition function. We introduce a formula derived in \cite{Jensen:2013rga} that determines Chern-Simons terms in the effective action after the circle reduction. We also derive the Cardy formula (\ref{eq89_general}) from the anomaly polynomial (\ref{eq95_anom}). In section \ref{section3}, we compute the Cardy free energies of general 6d (2,0) theories and some specific examples of (1,0) theories with flavor symmetries. Our Cardy formulas are checked by comparing with the superconformal indices whose closed forms are known. In section \ref{section4}, we find that the Cardy free energy is given by the equivariant integral of the thermal anomaly polynomial. In section \ref{section5}, we compute the entropy from the Cardy free energy. We show that the large \(N\) free energies of (2,0)  SCFTs exactly account for the entropy of BPS black holes in the dual \(AdS_7\). Also, we compute the asymptotic entropy in the Cardy limit and derive the bound (\ref{eq99_ineq}).

\section{Thermal derivative expansion}
\label{section2}
In this section, we review the derivative expansion of the thermal partition function of the even-dimensional QFT. We consider a QFT on \(\mathcal{M}^d=M^{d-1}\times S^1\) where \(d=2n\). \(M^{d-1}\) is \(d-1\)-dimensional compact manifold with \(\partial M^{d-1}=0\) and \(S^1\) is the Euclidean temporal circle which is fibered over \(M^{d-1}\). We will assume that the QFT has a global symmetry \(F\). The partition function takes the following definition,
\begin{align}
    Z(\beta,\omega_I,\Delta_i)=\text{Tr}\Bigr[ e^{-\beta E} \cdot \prod_{I=1}^{r_M} e^{-\omega_I J_I} \cdot \prod_{i=1}^{r_F} e^{-\Delta_i Q_i} \Bigr]
    \label{eq111_partition}
\end{align}
where \(E\) is the energy, \(J_I\)'s are the angular momenta of \(M^{d-1}\), and \(Q_i\)'s are the Cartan charges of the global symmetry \(F\). The partition function depends on the chemical potentials \(\beta\), \(\omega_I\), and \(\Delta_i\). The index \(i\) runs from \(1\) to \(r_F\) which is the rank of \(F\), and \(I\) runs from \(1\) to \(r_M\) which is a rank of the isometry of \(M^{d-1}\). \(\beta\) is the circumference of \(S^1\).
\par 
The thermal partition function can be computed from the path integral of dynamical fields. The chemical potentials are encoded in the background fields \(g_{\mu\nu}\) and \(A_\mu\). The background metric \(g_{\mu\nu}\) of \(\mathcal{M}^d\) depends on \(\omega_I\)'s, and the background gauge field \(A_\mu\) depends on \(\Delta_i\)'s. After integrating out all the dynamical fields, the partition function can be written as
\begin{align}
Z=\exp\Bigr[ -W(g_{\mu\nu},A_\mu)\Bigr]
\label{eq157_zero}
\end{align}
where \(W\) is the effective action of the background fields.
\par 
However, it is impossible to evaluate the effective action in a generic setting since the path integral cannot be performed exactly. Instead, it is helpful to consider the small circle limit \(\beta\ll \ell\) where \(\ell\) is a curvature radius of \(M^{d-1}\). In this limit, the dynamical fields can be reduced on \(M^{d-1}\) into zero-modes with zero \(S^1\) momentum and the Kaluza-Klein modes with non-zero \(S^1\) momentum. The KK modes have the mass proportional to \(\beta^{-1}\), and the path integral can be effectively evaluated as Gaussian integral. The zero-mode contribution to the effective action is subtle since they are light degrees on \(M^{d-1}\). However, in section \ref{section22}, we will introduce a certain scaling limit that the zero-mode contributions are suppressed.
\par 
One can canonically decompose the background fields on \(\mathcal{M}^d\) into \(S^1\) part and \(M^{d-1}\) part. For the metric \(g_{\mu\nu}\), it takes the following decomposition,
\begin{align}
    g_{\mu\nu}dx^\mu dx^\nu=e^{-2\Phi(x)}(d\tau+\mathbf{a} )+h_{ij} dx^i dx^j,\quad \tau\sim\tau+\beta,\quad (1\leq i,j\leq d-1)
    \label{eq163_reduction}
\end{align}
where \(\Phi\) is a dilaton field, \(\mathbf{a}=a_i dx^i\) is a graviphoton 1-form, and \(h_{ij}\) is a metric of \(M^{d-1}\). The background gauge field \(\mathbf{A}=A_i dx^i\) and the affine connection \(\mathbf{\Gamma}=(\Gamma^\mu{}_\nu)_ \rho dx^\rho\) are also decomposed as
\begin{align}
    \mathbf{A}=A_0(d\tau+\mathbf{a})+\hat{\mathbf{A}},\quad 
    \mathbf{\Gamma}=\Gamma_0(d\tau+\mathbf{a})+\hat{\mathbf{\Gamma}}
\end{align}
where \(A_0, \ \Gamma_0\) are temporal components of \(\mathbf{A}, \ \mathbf{\Gamma}\). Note that 1-forms \(\mathbf{a}\), \(\hat{\mathbf{A}}\), and \(\hat{\mathbf{\Gamma}}\) do not have a temporal component. 
\par 
The effective action \(W\) in (\ref{eq157_zero}) depends on the metric \(h_{ij}\), 1-forms \(\mathbf{a}\), \(\hat{\mathbf{A}}\), \(\hat{\mathbf{\Gamma}}\), and scalars \(A_0\), \(\Gamma_0\), \(\Phi\). Integrating out the KK modes yields the contribution to the effective action arranged as the form of an infinite series of the derivative expansion, which is the perturbative series of small \(\beta\). Due to the presence of the zero-modes, effective action can also have non-perturbative terms of small \(\beta\). The small \(\beta\) limit of the partition function is of great importance in our case since it is related to the `index point' where the partition function turns into the index \cite{Choi:2018hmj}. In section \ref{section21}, we will determine Chern-Simons terms in the effective action from the 't Hooft anomaly by following the works of Jensen, Loganayagam, and Yarom \cite{Jensen:2012kj,Jensen:2013kka,Jensen:2013rga}. In section {\ref{section22}}, we will explain the index limit of the partition function and show that the Cardy free energy can be determined from the Chern-Simons action only.

\subsection{Chern-Simons action and 't Hooft anomaly}
\label{section21}
The 't Hooft anomaly is the anomaly of global symmetries induced by chiral fields in even dimensions \cite{tHooft:1977nqb}. The naive path integral yields the effective action of the background fields which suffers from the following 't Hooft anomaly,
\begin{align}
\delta  W_\text{cons}= -2\pi i \int_{\mathcal{M}^d} I_{d}[g_{\mu\nu},A_\mu]
\label{eq145_variation}
\end{align}
where \(\delta\) collectively denotes the background gauge and the diffeomorphism transformation. We will call the effective action in (\ref{eq145_variation}) as the `consistent action' since currents obtained from \(W_\text{cons}\) satisfy the Wess-Zumino consistency condition \cite{Wess:1971yu}. The \(d\)-form anomaly \(I_d\) is determined from the \(d+2\)-form anomaly polynomial \(P_{d+2}\) through the following anomaly descent mechanism,
\begin{align}
dI_{d}=\delta I^\text{CS}_{d+1},\quad dI^\text{CS}_{d+1}=P_{d+2}
\label{eq_descent}
\end{align}
where \(I^\text{CS}_{d+1}\) is a Chern-Simons (CS) \(d+1\)-form made of \(\mathbf{A}\) and \(\mathbf\Gamma\). See \cite{Bilal:2008qx} for a detailed review of the anomalies.
\par 
In this subsection, we focus on the CS terms in the effective action. In \(W_\text{cons}\), there are two types of the CS actions: the gauge/diffeomorphism invariant CS action \(W_\text{ICS}\) and the non-invariant CS action \(W_\text{NCS}\). \(W_\text{ICS}\) consists CS forms made of 1-forms \(\hat{\mathbf{a}},\ \hat{\mathbf{A}}\) and \(\hat{\mathbf{\Gamma}}\) with properly quantized coefficients, and therefore they are invariant under the background gauge/diffeomorphism transformation. On the other hand, \(W_\text{NCS}\) takes the form of normal CS terms but multiplied with the scalars \(A_0\), \(\Gamma_0\) and therefore it is the only gauge/diffeomorphism non-invariant part in the effective action. As a result, \(W_\text{NCS}\) is directly determined from the 't Hooft anomaly (\ref{eq145_variation}) as follows \cite{Jensen:2013rga},
\begin{align}
 W_\text{NCS}=-2\pi i \int_{\mathcal{M}^d} {dt+\mathbf{a}\over d\mathbf{a} }\Bigr[ I^\text{CS}_{d+1}-\hat{I}^\text{CS}_{d+1} \Bigr].
 \label{eq_wncs}
\end{align}
The hatted form \(\hat{I}^\text{CS}_{d+1}\) is defined from \(I^\text{CS}_{d+1}\) by the following replacement,
\begin{align}
\hat{I}^\text{CS}_{d+1}=I^\text{CS}_{d+1}\Bigr(\mathbf{A}\to\hat{\mathbf{A}}, \ \mathbf{\Gamma}\to\hat{\mathbf{\Gamma}} \Bigr).
\label{eq177_hat}
\end{align}
Throughout this paper, we will define the hatted notation as replacing 1-forms to their hatted forms as (\ref{eq177_hat}). The two-form \(d\mathbf{a}\) in the denominator means the following operation,
\begin{align}
{1\over d\mathbf{a}}\sum_{k=1} c_k (d\mathbf{a})^k=\sum_{k=1} c_k (d\mathbf{a})^{k-1}
\end{align}
which is well-defined since \((dt+\mathbf{a})(I^\text{CS}_{d+1}-\hat{I}^\text{CS}_{d+1})\) can be written as a finite power series of \(d\mathbf{a}\) which starts from \((d\mathbf{a})^1\). Then it is straightforward to check that the variation of \(W_\text{NCS}\) generates the 't Hooft anomaly of the consistent action (\ref{eq145_variation}) as follows,
\begin{align}
    \delta W_\text{NCS}
    &=-2\pi i \int_{\mathcal{M}^d} {dt+\mathbf{a}\over d\mathbf{a} }\Bigr[ dI_{d}-d\hat{I}_{d} \Bigr]
    =-2\pi i \int_{\mathcal{M}^d}\Bigr[ I_{d}-\hat{I}_{d} \Bigr]
    =-2\pi i \int_{\mathcal{M}^d} I_{d}.
    \label{eq190_wncs}
\end{align}
At the second equality of (\ref{eq190_wncs}), we used that \(\partial \mathcal{M}^d=0\) and \(d{dt+\mathbf{a}\over d\mathbf{a}}=1\). Although \({dt+\mathbf{a}\over d\mathbf{a}}\) is not a properly defined differential form, it can be regarded as an \(-1\)-form and \(d{dt+\mathbf{a}\over d\mathbf{a}}=1\) means the following property,
\begin{align}
d\Bigr[{dt+\mathbf{a}\over d\mathbf{a}} \sum_{k=1} c_k (d\mathbf{a})^k \Bigr]
&=\sum_{k=1} c_k (d\mathbf{a})^k -{dt+\mathbf{a} \over d\mathbf{a}}d\Bigr[ \sum_{k=1} c_k (d\mathbf{a})^k  \Bigr].
\end{align}
At the third equality of (\ref{eq190_wncs}), we used that \(\hat{I}_d\) does not contain the \(d\tau\) component and vanish when integrating over \(S^1\).
\par 
By varying the consistent effective action with respect to the background fields, we obtain the consistent current \(J^\mu_\text{cons}={\delta W_\text{cons}\over \delta A_\mu}\) and the consistent energy-momentum tensor \(T^{\mu\nu}_\text{cons}={\delta W_\text{cons}\over \delta g_{\mu\nu}}\). However, they do not have the covariant form under the gauge/diffeomorphism transformation. This non-covariance can be cured by adding the local terms to \(J^\mu_\text{cons}\) and \(T^{\mu\nu}_\text{cons}\) which are called the Bardeen-Zumino polynomials \cite{Bardeen:1984pm}. The covariant current and the energy-momentum tensor can be obtained by varying the `covariant action' \(W_\text{cov}\) which is defined as follows,
\begin{align}
W_\text{cov}=W_\text{cons}+W_\text{bulk},\quad W_\text{bulk}=2\pi i \int_{\mathcal{N}^{d+1}} I^\text{CS}_{d+1}
\label{eq158_BZ}
\end{align}
where \(\mathcal{N}^{d+1}\) is the bulk \(d+1\)-dimensional manifold whose boundary is given by \(\mathcal{M}^d\). More precisely, we set \(\mathcal{N}^{d+1}=S^1\times \mathcal{N}^d\) where \(\partial N^d=M^{d-1}\). In the covariant action, \(W_\text{NCS}+W_\text{bulk}\) can be written as the following expression \cite{Jensen:2013rga},
\begin{align}
 W_\text{NCS}+W_\text{bulk}=2\pi i \int_{\mathcal{N}^{d+1}} {dt+\mathbf{a}\over d\mathbf{a} }\Bigr[ P_{d+2}-\hat{P}_{d+2} \Bigr]
 \label{eq204_bzncs}
\end{align}
which can be proven from the following identity,
\begin{align}
{dt+\mathbf{a}\over d\mathbf{a}}\Bigr(P^\text{CS}_{d+1}-\hat{P}^\text{CS}_{d+1} \Bigr)=\Bigr(I^\text{CS}_{d+1}-\hat{I}^\text{CS}_{d+1} \Bigr)-d\Bigr[{dt+\mathbf{a}\over d\mathbf{a}}\Bigr(I^\text{CS}_{d+1}-\hat{I}^\text{CS}_{d+1} \Bigr)\Bigr].
\end{align}
After integrating over \(S^1\), the contribution from \(\hat{I}^\text{CS}_{d+1}\) vanishes due to its absence of the temporal component, and we obtain the desired result (\ref{eq204_bzncs}).
\par 
Until now, we have determined \(W_\text{NCS}\) and \(W_\text{bulk}\) which are not invariant under the background gauge/diffeomorphism transformation. However, the invariant CS action \(W_\text{ICS}\) is not anomalous under the background gauge/diffeomorphism transformation. As a result, \(W_\text{ICS}\) is not directly related to the anomalous transformation of the effective action (\ref{eq145_variation}).  However, in \cite{Jensen:2012kj,Jensen:2013kka,Jensen:2013rga}, it was pointed out that one can impose another consistency condition called a `consistency with the Euclidean vacuum' to determine \(W_\text{ICS}\). Here, we introduce the condition and briefly sketch the derivation presented in \cite{Jensen:2013rga}. 
\par 
Let us consider a \(d\)-dimensional manifold \(\mathcal{M}^d=\mathbb{R}^{2,*}\times M^{d-2}\) where \(\mathbb{R}^{2,*}\) is the two-dimensional Euclidean plane with the origin removed. The plane \(\mathbb{R}^{2,*}\) is fibered over the \(d-2\)-dimensional manifold \(M^{d-2}\), and the fibration depends on the graviphoton \(\mathbf{a}\). The metric of \(\mathbb{R}^{2,*}\) is given by \(ds^2_{\mathbb{R}^{2,*}}=r^2 d\theta^2+dr^2\) with \(\theta\sim\theta+2\pi\) periodicity. We formally interpret \(\theta\) as the Euclidean time \(\tau\) by identifying \(\tau={\beta\over 2\pi}\theta\), thus giving the metric \(ds^2_{\mathbb{R}^{2,*} }={4\pi^2 r^2\over \beta^2}d\tau^2+dr^2\). The total metric of \(\mathcal{M}^d\) is then given by follows,
\begin{align}
ds^2={4\pi^2 r^2\over \beta^2}(d\tau+\mathbf{a})^2+dr^2+ds^2_{M^{d-2}}.
\end{align}
Now, we turn on the specific background field configuration of \(\mathbf{a}\) and \(\hat{\mathbf{A}}\) which are non-zero only on \(M^{d-2}\). Let us consider an infinitesimal metric perturbation on \(\mathbb{R}^{2,*}\) as \(\delta g_{\tau r}d\tau dr\). The consistency with the Euclidean vacuum requires the following condition,
\begin{align}
 T^{\tau r}_\text{cov} ={\partial W_\text{cov}\over \delta g_{\tau r}}=0.
\label{eq221_consistency}
\end{align}
Since the graviphoton \(\mathbf{a}\) and the background gauge field \(\hat{\mathbf{A}}\) are only turned on \(M^{d-2}\), the partition function should reproduce correlation functions of the Euclidean vacuum on \(\mathbb{R}^{2,*}\). As a result, the rotational invariance of the Euclidean vacuum is equivalent to the consistency condition (\ref{eq221_consistency}). In this geometric setting, it should be noted that the derivative expansion of \(W_\text{cov}\) is generally ineffective since the angular circle of \(\mathbb{R}^{2,*}\) becomes large if we go far from the origin. In this case, the infinite series of the derivative expansion has to be resumed, as well as non-perturbative terms, in order to compute the effective action or the generic correlation functions. However, in \cite{Jensen:2013rga}, it was argued that one can construct certain background field configurations such that only finite number of terms in \(W_\text{bulk}\), \(W_\text{NCS}\), and \(W_\text{ICS}\) contribute to \(T^{\tau r}_\text{cov}\) among infinitely many terms in the effective action. In such cases, \(T^{\tau r}\) can be exactly computed regardless of the validity of the derivative expansion. As a result, the consistency condition (\ref{eq221_consistency}) can be used to determine \(W_\text{ICS}\) since \(W_\text{bulk}\) and \(W_\text{NCS}\) are already determined as (\ref{eq204_bzncs}).
\par 
As a conclusion, together with the results shown in (\ref{eq_wncs}) and (\ref{eq158_BZ}), the three types of CS actions in \(W_\text{cov}\) take the following expression,
\begin{align}
&W_\text{bulk}=2\pi i \int_{\mathcal{N}^{d+1} } I^\text{CS}_{d+1}
\nonumber \\
&W_\text{NCS}=-2\pi i \int_{\mathcal{M}^{d}} {dt+\mathbf{a}\over d\mathbf{a}}\Bigr[I^\text{CS}_{d+1}-\hat{I}^\text{CS}_{d+1} \Bigr]
\nonumber \\
&W_\text{ICS}=-2\pi i \int_{\mathcal{M}^{d}} {dt+\mathbf{a}\over d\mathbf{a}}\Bigr[\Delta I^\text{CS}_{d+1}-\Delta \hat{I}^\text{CS}_{d+1} \Bigr]_{\mathbf{ A}^T=0}.
\label{eq209_three}
\end{align}
Here, \(\Delta I^\text{CS}_{d+1}\) is another CS \(d+1\)-form which will be explained shortly, and \(\mathbf{A}^T\) is a fictitious \(U(1)\) gauge field defined as follows,
\begin{align}
\mathbf{A}^T={2\pi i \over \beta}(d\tau+\mathbf{a})+\hat{\mathbf{A}}^T.
\label{eq234_fic}
\end{align}
Using \(\mathbf{A}^T\), we define a `thermal anomaly polynomial' from the ordinary anomaly polynomial by the following replacement rule \cite{Jensen:2013rga},
\begin{align}
P^T_{d+2}=P_{d+2}\Bigr(p_k(T)\to p_k(T)-{\mathbf{F}_T^2\over 4\pi^2}  p_{k-1}(T) \Bigr)
\label{eq273_repla}
\end{align}
where \(p_k(T)\) is the \(k\)'th Pontryagin class of the tangent bundle. Then, \(\Delta I^\text{CS}_{d+1}\) is defined as
\begin{align}
P^T_{d+2}=dI_{d+1}^T \quad \text{where}\quad  I_{d+1}^T=I^\text{CS}_{d+1}+\Delta I^\text{CS}_{d+1}.
\end{align}
Note that the fictitious field \(\mathbf{A}^T\) will be turned off at the end of the computation, but its effect on \(\Delta\hat{I}^\text{CS}_{d+1}\) survives since \(\hat{\mathbf{A}}^T\) is non-zero. The sum of the three actions in (\ref{eq209_three}) can be written as the following simple form,
\begin{align}
 W_\text{bulk}+W_\text{NCS}+W_\text{ICS}=2\pi i \int_{\mathcal{N}^{d+1}} {dt+\mathbf{a}\over d\mathbf{a} }\Bigr[ P^T_{d+2}-\hat{P}^T_{d+2} \Bigr]_{\mathbf{A}^T=0}.
\label{eq229_cs}
\end{align}
The dimensional reduction of (\ref{eq229_cs}) gives the CS forms on \(M^{d-1}\) which are the only CS terms in \(W_\text{cov}\). Therefore, all CS terms in the covariant action can be computed from the anomaly polynomial of the theory.
\par 
For the quantized charges, the chemical potentials in (\ref{eq111_partition}) have the periodicity along the imaginary direction, and their periodic shift is related to the large gauge transformation on \(S^1\). If we properly regulate the partition function to preserve the periodicity of the chemical potentials, it is natural to obtain the covariant action as the effective action in (\ref{eq157_zero}). The reason is that the covariant action itself is invariant under the background gauge/diffeomorphism transformation even when the 't Hooft anomaly is non-zero. Therefore, in the rest of this paper, we will use \(W_\text{cov}\) to compute the partition function defined in (\ref{eq111_partition}) and (\ref{eq157_zero}) to ensure the periodicity of the chemical potentials. As a related matter, the presence of the Casimir energy prefactor can break the periodicity of the chemical potential. Therefore, we speculate that the covariant action is related to the spectral/state-counting part of the superconformal index. In section \ref{section3}, it will be shown that our Cardy formula gives the Cardy free energy of the superconformal index without the Casimir energy prefactor.
\par 
Aside from the CS action, the effective action receives two other types of contributions, which we will denote \(W_\text{non-CS}\) and \(W_\text{zero}\). \(W_\text{non-CS}\) is the non-CS terms in the derivative expansion, and \(W_\text{zero}\) is the terms that are not captured in the derivative expansion due to the contribution of the dynamical zero-modes. As a result, the partition function (\ref{eq111_partition}) can be written as
\begin{align}
Z=\exp\Bigr[-W_\text{CS}-W_\text{non-CS}-W_\text{zero} \Bigr]
\label{eq255_three}
\end{align}
where \(W_\text{CS}\equiv W_\text{ICS}+W_\text{NCS}+W_\text{bulk}\) is the CS action determined in (\ref{eq229_cs}). Without evaluating \(W_\text{non-CS}\) and \(W_\text{zero}\), the partition function can not be precisely obtained. However, in the following subsection, we will show that in the certain scaling limit where the partition function turns into the index, the evaluation of \(W_\text{CS}\) is sufficient to determine the behavior of the index in the Cardy limit.

\subsection{6d index from Chern-Simons action}
\label{section22}
In this section, we focus on 6d superconformal field theories with \(SU(2)_R\) R-symmetry on \(S^5\times S^1\). We will show that the modified index can be obtained from the small \(S^1\) limit of the partition function. If we further take the Cardy limit \((|\omega_I|\ll 1)\) after taking the small \(S^1\) limit, the leading term and a few subleading terms of the free energy can be determined from the CS action on \(S^5\).
\par 
The 6d modified superconformal index (\ref{eq65_modind}) counts BPS states on \(S^5\), and takes the following definition,
\begin{align}
I=\text{Tr}\Bigr[e^{-\Delta_R Q_R}\cdot e^{-\omega_1 J_1-\omega_2 J_2-\omega_3 J_3} \Bigr]
\label{eq238_index}
\end{align}
where the chemical potentials are constrained by \(\Delta_R-\omega_1-\omega_2-\omega_3=2\pi i\) and  the other flavor symmetry chemical potential is turned off (\(m_i=0\)). We turn off the other global symmetries for simplicity. The partition function of the same 6d theory takes the following definition,
\begin{align}
Z=\text{Tr}\Bigr[ e^{-\beta E}\cdot e^{-\Delta_R Q_R} \cdot e^{-\omega_1 J_1-\omega_2 J_2-\omega_3 J_3}\Bigr].
\label{eq288_part}
\end{align}
For the partition function, the chemical potentials are all independent, and it counts all states on \(S^5\).  However, one can turn the partition function to the index by imposing the chemical potential constraint \(\Delta_R-\omega_1-\omega_2-\omega_3=2\pi i\) and taking \(\beta/r\to 0 \) limit as follows,
\begin{align}
I=\lim_{\beta/ r \to 0} Z \quad \text{at}\quad \Delta_R-\omega_1-\omega_2-\omega_3=2\pi i
\label{eq293_ind}
\end{align}
where \(r\) is the radius of \(S^5\).
\par 
The original Cardy formula of 2d CFT \cite{Cardy:1986ie} counts the degeneracy of the states in a large spatial momentum limit, which is called the Cardy limit. The higher dimensional generalization of the Cardy limit can be understood as the limit where \(J_I\)'s are large. In the canonical ensemble, such a limit can be obtained by taking \(|\omega_I|\ll 1\) \cite{Choi:2018hmj}. The Cardy limit \(|\omega_I|\ll 1\) of the index \(I\) can be studied from the partition function \(Z\) by taking the \(\beta\ll r\) limit first, and then taking \(|\omega_I|\ll 1\) limit later. It can be accomplished by considering the following scaling limit,
\begin{align}
{\beta \over r}\ll |\omega_{1,2,3}| \ll 1.
\label{eq296_scaling}
\end{align}
Now, let us evaluate the various terms of the effective action in the above scaling limit (\ref{eq296_scaling}).
\par 
With the chemical potentials turned on as (\ref{eq288_part}), the background metric on \(S^5\times S^1\) is given as follows,
\begin{align}
g_{\mu\nu}dx^\mu dx^\nu&=r^2\Bigr[d\theta_1^2+\sin^2 \theta_1 d\theta_2^2+\sum_{i=1}^3 n_i^2 \Bigr(d\phi_i-{i\omega_i\over \beta}d\tau\Bigr)^2 \Bigr]+d\tau^2
\nonumber \\
n_1&=\cos\theta_1, \quad n_2=\sin\theta_1 \cos\theta_2,\quad n_3=\sin\theta_1 \sin\theta_2
\label{eq284_metric}
\end{align}
with temporal periodicity \(\tau\sim\tau+\beta\). The background \(SU(2)_R\) R-symmetry gauge field \(\mathbf{A}_{SU(2)}\) is turned on as
\begin{align}
\mathbf{A}_{SU(2)} = \mathbf{A}\cdot T_{SU(2)},\quad \mathbf{A}={\Delta_R\over \beta}d\tau
\label{eq290_AI}
\end{align}
where \(T_{SU(2)}\) is the Cartan generator of \(SU(2)\) normalized as \(\text{tr}_\textbf{adj}[T_{SU(2)}^2]=2\), and \(\mathbf{A}\) is the \(U(1)\) Cartan gauge field. One may worry that due to \(\beta\) in the denominators of the background fields, the effective action might be singular in our scaling limit. However, we will show that effective action is well-defined in our scaling limit. A similar setting was considered in \cite{Choi:2018hmj}.
\par 
As introduced in (\ref{eq163_reduction}), the metric and the gauge fields are reduced on \(S^5\) as follows,
\begin{align}
g_{\mu\nu}dx^\nu dx^\nu&=e^{-2\Phi} (d\tau+\mathbf{a})^2+h_{ij}dx^i dx^j
\nonumber \\
h_{ij}dx^i dx^j&=r^2 \Bigr( d\theta_1^2+\sin^2\theta_1 d\theta_2^2+n_i^2 d\phi_i^2+{r^2\over \beta^2}{(\sum_i \omega_i n_i^2 d\phi_i)^2\over 1-{r^2\over \beta^2}\sum_i n_i^2 \omega_i^2} \Bigr)
\label{eq229_metric}
\end{align}
where \(h_{ij}\) is a metric of the squashed \(S^5\). We will temporarily assume that \(\omega_i\)'s are purely imaginary so that the metric (\ref{eq229_metric}) is well defined. The extension to the complex \(\omega_i\)'s can be understood as the analytic continuation. The dilaton \(\Phi\) and the graviphoton \(\mathbf{a}\) are
\begin{align}
e^{-2\Phi}=1-{r^2\over \beta^2}\sum_i n_i^2 \omega_i^2,\quad 
\mathbf{a}=-i{r^2 \sum_i \omega_i n_i^2 d\phi_i\over \beta \Bigr(1-{r^2\over \beta^2}\sum_i n_i^2 \omega_i^2\Bigr)}.
\label{eq318_gravi}
\end{align}
The background gauge field and the affine connection are
\begin{align}
    \mathbf{A}=A_0(d\tau+\mathbf{a})+\hat{\mathbf{A}},\quad 
    \mathbf{\Gamma}=\Gamma_0(d\tau+\mathbf{a})+\hat{\mathbf{\Gamma}}.
\end{align}
We should keep in mind that, due to (\ref{eq290_AI}), we should plug \(A_0={\Delta_R\over \beta}\) and \(\hat{\mathbf{A}}=-A_0 \mathbf{a}\) after the computation. Now, let us evaluate the three contributions \(W_\text{CS}\), \(W_\text{non-CS}\), and \(W_\text{zero}\) in the effective action (\ref{eq255_three}) one by one.
\par 
First, we consider the CS terms \(W_\text{CS}\) on \(S^5\). The explicit structure of the CS action is determined from the thermal anomaly polynomial as (\ref{eq229_cs}). Let us consider the most general form of the 6d anomaly polynomial with \(SU(2)\) R-symmetry and the tangent bundle as follows,
\begin{align}
P_8={1\over 4!}\Bigr(\mathfrak{a} \cdot c_2(R)^2+\mathfrak{b}\cdot c_2(R) p_1(T)+\mathfrak{c} \cdot p_1(T)^2+\mathfrak{d} \cdot p_2(T)  \Bigr)
\label{eq319_anom}
\end{align}
where \(\mathfrak{a},\mathfrak{b},\mathfrak{c}\) and \(\mathfrak{d}\) are the 't Hooft anomaly coefficients. The second Chern class of \(SU(2)\) R-symmetry is defined as
\begin{align}
c_2(R)={1\over 4(2\pi)^2}\text{Tr}[\mathbf{F}_{SU(2)}^2]={1\over 4(2\pi)^2}\mathbf{F}^2
\label{eq329_c2}
\end{align}
where the capital trace is defined as \(\text{Tr}\equiv (h_G^\vee)^{-1} \text{tr}_\textbf{adj}\) and \(\mathbf{F}=d\mathbf{A}\). Note that the dual Coxeter number of \(SU(2)\) is \(h_G^\vee=2\). The first and the second Pontryagin classes of the tangent bundle are
\begin{align}
p_1(T)=-{1\over 2(2\pi)^2}\text{tr}[\mathbf{R}^2],\quad 
p_2(T)={1\over (2\pi)^4}\Bigr(-{1\over 4}\text{tr}[\mathbf{R}^4]+{1\over 8}(\text{tr}[\mathbf{R}^2])^2 \Bigr)
\label{eq335_p12}
\end{align}
where \(\mathbf{R}\) is a curvature two-form of 6d manifold. After applying the replacement rule (\ref{eq273_repla}), the thermal anomaly polynomial is
\begin{align}
P_8^T={1\over 4!}\Bigr[\mathfrak{a}\cdot c_2(R)^2+\mathfrak{b}\cdot c_2(R) \Bigr( p_1(T) -{\mathbf{F}_T^2\over 4\pi^2}\Bigr)+\mathfrak{c} \cdot \Bigr(p_1(T)-{\mathbf{F}_T^2\over 4\pi^2}\Bigr)^2+\mathfrak{d}\cdot \Bigr(p_2(T)-{\mathbf{F}_T^2\over 4\pi^2} p_1(T)\Bigr)  \Bigr]
\end{align}
where \(\mathbf{F}_T\) is the field strength of a fictitious \(U(1)\) gauge field defined in (\ref{eq234_fic}). The thermal anomaly polynomial can be rearranged as follows,
\begin{align}
P_8^T&={\mathfrak{a}\over 6144\pi^4} \mathbf{F}^4-{\mathfrak{b}\over 1536\pi^4}\mathbf{F}^2 \mathbf{F}_T^2+{\mathfrak{c}\over 384\pi^4} \mathbf{F}_T^4 
& &\to\quad \text{gauge anomaly}
\nonumber \\
&-{\mathfrak{b}\over 3072\pi^4} \text{tr}[\mathbf{R}^2] \mathbf{F}^2
+{2\mathfrak{c}+\mathfrak{d}\over 768\pi^4} \text{tr}[\mathbf{R}^2] \mathbf{F}_T^2
 & &\to\quad \text{mixed anomaly}
\nonumber \\
&-{\mathfrak{d}\over 1536\pi^4}\text{tr}[\mathbf{R}^4]
+{2\mathfrak{c}+\mathfrak{d}\over 3072\pi^4}\text{tr}[\mathbf{R}^2]^2
 & &\to\quad \text{gravitational anomaly}.
 \label{eq327_anom}
\end{align}
In (\ref{eq327_anom}), we divide the anomalies into the three types: gauge, mixed, and gravitational anomalies. The CS action is then given by
\begin{align}
W_\text{CS}=2\pi i \int_{\mathcal{N}^{7}} {d\tau+\mathbf{a}\over d\mathbf{a} }\Bigr[ P^T_{8}-\hat{P}^T_{8} \Bigr]_{\mathbf{A}^T=0}
\label{eq331_CS}
\end{align}
where \(\mathcal{N}^7=B^6\times S^1\) with \(\partial B^6=S^5\) such as a six-dimensional solid ball. The two-forms can be written in terms of hatted two-forms as follows.
\begin{align}
&\mathbf{F}=d\mathbf{A}=\hat{\mathbf{F}}+A_0 d\mathbf{a}+\mathcal{O}(d\tau+\mathbf{a})
\nonumber \\
&\mathbf{R}=d\mathbf{\Gamma}+\mathbf{\Gamma}^2=\hat{\mathbf{R}}+\Gamma_0 d\mathbf{a} +\mathcal{O}(d\tau+\mathbf{a})
\nonumber \\
&\mathbf{F}_T=\hat{\mathbf{F}}_T+{2\pi i \over\beta}d\mathbf{a}.
\end{align}
Then, the thermal anomaly polynomial \(P_8^T\) can be written as
\begin{align}
P_8^T=P_8^T\Bigr(\mathbf{F}\to\hat{\mathbf{F}}+A_0 d\mathbf{a}, \mathbf{R}\to\hat{\mathbf{R}}+\Gamma_0 d\mathbf{a},\mathbf{F}_T\to \hat{\mathbf{F}}_T+{2\pi i\over \beta}d\mathbf{a} \Bigr)+\mathcal{O}(d\tau+\mathbf{a}).
\end{align}
The terms of order \(\mathcal{O}(d\tau+\mathbf{a})\) do not contribute to the CS action since they vanish when evaluating (\ref{eq331_CS}).
\par 
Now, let us compute the CS action from the thermal anomaly polynomial (\ref{eq327_anom}). We first consider the gauge anomaly which gives the gauge CS action as follows,
\begin{align}
W_\text{gauge-CS}=2\pi i \int_{\mathcal{N}^7} {d\tau+\mathbf{a}\over d\mathbf{a}}
\Bigr[& {\mathfrak{a}\over 6144\pi^4} \Bigr( (\hat{\mathbf{F}}+A_0 d\mathbf{a})^4-\hat{\mathbf{F}}^4 \Bigr)
\nonumber \\
-&{\mathfrak{b}\over 1536\pi^4}\Bigr( (\hat{\mathbf{F}}+A_0 d\mathbf{a})^2 (\hat{\mathbf{F}}_T+{2\pi i \over \beta}d\mathbf{a})^2-\hat{\mathbf{F}}^2\hat{\mathbf{F}}_T^2 \Bigr)
\nonumber \\
+&{\mathfrak{c}\over 384\pi^4}\Bigr( (\hat{\mathbf{F}}_T+{2\pi i \over \beta}d\mathbf{a})^4 \Bigr)-\hat{\mathbf{F}}_T^4 \Bigr]_{\mathbf{A}_T=0}.
\end{align}
The integral is defined on the \(\mathcal{N}^7=S^1\times B^6\), but it can be reduced on the boundary since \(A_0\) is a constant. After integrating over \(S^1\), we obtain,
\begin{align}
&W_\text{gauge-CS}
\nonumber \\
&=2\pi i \beta \int_{S^5}\Bigr[ {\mathfrak{a}\over 6144\pi^4}  \Bigr(A_0^4 \mathbf{a} d\mathbf{a} d\mathbf{a}+4 A_0^3 \hat{\mathbf{A}}d\mathbf{a}d\mathbf{a}+6A_0^2 \mathbf{a} d\hat{\mathbf{A}}d\hat{\mathbf{A}}+4A_0 \hat{\mathbf{A}}d\hat{\mathbf{A}} d\hat{\mathbf{A}} \Bigr)
\nonumber \\
&\hspace{2cm}-{\mathfrak{b}\over 384 \pi^2 \beta^2} \mathbf{a}d\hat{\mathbf{A}} d\hat{\mathbf{A}}
-{\mathfrak{c} \over 24\pi^2 \beta^4}\mathbf{a}d\mathbf{a}d\mathbf{a}\Bigr]
\nonumber \\
&=2\pi i  \int_{S^5}\Bigr(-{\mathfrak{a}\over 6144\pi^4}  (\beta A_0)^4 -{\mathfrak{b}\over 384 \pi^2 } (\beta A_0)^2 -{\mathfrak{c} \over 24}\Bigr)
\cdot {1\over \beta^3}\mathbf{a}d\mathbf{a}d\mathbf{a}
\end{align} 
where we used \(\hat{\mathbf{A}}=-A_0 \mathbf{a}\) in the second equality. By a similar way, we can obtain the mixed CS action as follows,
\begin{align}
W_\text{mixed-CS}=2\pi i  \int_{S^5}
\Bigr({\mathfrak{b}\over 3072\pi^4} (\beta A_0)^2 
+{2\mathfrak{c}+\mathfrak{d}\over 192\pi^2 } \Bigr) \cdot {1\over \beta}\mathbf{a} \text{tr}[\mathbf{\hat{R} }^2].
\end{align}
The gravitational CS action is much more complicated, and it is unclear whether its form is uniquely fixed on \(S^5\). The main complication is that \(\Gamma_0\) is not a constant unlike to \(A_0\). Here, we list a few sample terms that can appear in the gravitational CS action as follows,
\begin{align}
W_\text{grav-CS}\supset \beta \int_{S^5}\text{tr}[\Gamma_0^4]\mathbf{a}d\mathbf{a}d\mathbf{a},\quad 
\beta \int_{S^5}\text{tr}[\Gamma_0^2]^2\mathbf{a}d\mathbf{a}d\mathbf{a},...
\end{align}
where many other complicated terms are omitted. However, we do not need to evaluate \(W_\text{gravi-CS}\) since they are subleading in our scaling limit, as will be discussed in the next paragraph.
\par 
In our scaling limit, those three types of CS terms have different scaling behavior. The gauge CS terms scales of order \(\mathcal{O}(({\beta\over r})^0 \omega^{-3})\), the mixed CS terms scales of order \(\mathcal{O}( ({\beta\over r})^0 \omega^{-1})\), and the gravitational CS terms scales of order \(\mathcal{O}( ({\beta\over r})^0 \omega \log \omega )\). For the gauge CS action, the corresponding CS term gives
\begin{align}
{1\over \beta^3}\int_{S^5} \mathbf{a} d\mathbf{a} d\mathbf{a}&=i{(2\pi)^3   \omega_1 \omega_2 \omega_3\over \Bigr( \omega_1^2-{\beta^2\over r^2} \Bigr)\Bigr(\omega_2^2- {\beta^2\over r^2} \Bigr)\Bigr( \omega_3^2-{\beta^2\over r^2} \Bigr)}={8i\pi^3 \over \omega_1 \omega_2 \omega_3}\Bigr[1+\mathcal{O}\Bigr( ({ \beta\over r\omega })^2\Bigr) \Bigr].
\label{eq329_CS}
\end{align}
For the mixed CS action, the corresponding CS term gives
\begin{align}
{1\over\beta}\int_{S^5} \mathbf{a}  \text{tr}[ \hat{\mathbf{R}}^2 ]&=2i{(2\pi)^3  \omega_1 \omega_2 \omega_3  (\omega_1^2+\omega_2^2+\omega_3^2)\over \Bigr( \omega_1^2-{\beta^2\over r^2} \Bigr)\Bigr(\omega_2^2- {\beta^2\over r^2} \Bigr)\Bigr( \omega_3^2-{\beta^2\over r^2} \Bigr) }={16i\pi^3  (\omega_1^2+\omega_2^2+\omega_3^2)\over \omega_1 \omega_2 \omega_3}\Bigr[1+\mathcal{O}\Bigr( ({ \beta\over r \omega })^2\Bigr) \Bigr].
\label{eq367_mixed}
\end{align}
For the gravitational CS action, we present a contribution from a single term as an example as follows,

\begin{align}
&\beta \int_{S^5} \text{tr}[\Gamma_0^4] \mathbf{a}d\mathbf{a}d\mathbf{a}=32 i \pi^3
\Bigr({\omega_1^4+\omega_2^4+\omega_3^4\over 2\omega_1 \omega_2 \omega_3} 
+{2\omega_1 \omega_2 \omega_3 (3\omega_1^2+\omega_2^2+\omega_3^2)
\over (\omega_1^2-\omega_2^2)(\omega_1^2-\omega_3^2) } \log \omega_1
\nonumber \\
&+{2\omega_1 \omega_2 \omega_3 (\omega_1^2+3\omega_2^2+\omega_3^2)
\over (\omega_2^2-\omega_1^2)(\omega_2^2-\omega_3^2) } \log \omega_2
+{2\omega_1 \omega_2 \omega_3 (\omega_1^2+\omega_2^2+3\omega_3^2)
\over (\omega_3^2-\omega_1^2)(\omega_3^2-\omega_2^2) } \log \omega_3 \Bigr)
\times \Bigr[1+\mathcal{O}\Bigr( ({ \beta\over r\omega })^2\Bigr) \Bigr].
\label{eq355_gravi}
\end{align}
In the constant dilaton background where \(\omega\equiv \omega_{1,2,3}\), we checked that the other gravitational CS terms are also of order \(\mathcal{O}(\omega \log \omega)\) or lower, which are subleading than the other CS terms. Moreover, as we will argue shortly, the zero-mode contribution \(W_\text{zero}\) is considered to be of order \(\mathcal{O}(\log \omega)\) which is more dominant than the gravitational CS terms. Since the zero-mode contribution cannot be determined in the anomaly-based approach, we will ignore the gravitational CS terms, and keep the only gauge and mixed CS terms to compute the Cardy free energy. From (\ref{eq329_CS}) and (\ref{eq367_mixed}), we obtain the following results for the anomaly polynomial given in (\ref{eq319_anom}),
\begin{align}
W_\text{gauge-CS}&=\Bigr( { \mathfrak{a}\over 384}\Delta_R^4+{\mathfrak{b} \pi^2\over 24}\Delta_R^2+{2\mathfrak{c}\pi^4\over 3} \Bigr){1\over \omega_1 \omega_2 \omega_3}+\mathcal{O}\Bigr( ({ \beta\over r\omega })^2\Bigr) 
\nonumber \\
W_\text{mixed-CS}&=-\Bigr({\mathfrak{b}\over 96}\Delta_R^2+{(2\mathfrak{c}+\mathfrak{d} )\pi^2\over 6} \Bigr){\omega_1^2+\omega_2^2+\omega_3^2\over \omega_1 \omega_2 \omega_3}+\mathcal{O}\Bigr( ({ \beta\over r\omega })^2\Bigr).
\end{align}
\par 
Now, let us discuss the non-CS terms in the derivative expansion. In order to simplify our discussion, we shall consider the constant dilaton background \(\omega\equiv \omega_{1,2,3}\) when evaluating non-CS terms. The generic non-CS Lagrangian can be constructed from the graviphoton field strength  \(f_{ij}=\beta^{-1} (\partial_i a_j-\partial_j a_i)\), the gauge field strength \(F_{ij}=\partial_i \hat{A}_j-\partial_j \hat{A}_i\), the curvature \(\hat{R}^i{}_{jkl}\), and the two temporal scalars \(\beta A_0\), \(\beta \Gamma^i{}_{j0}\). The exceeding number of the lower indices should be contracted with the metric \( (\beta e^{-\Phi})^2 h^{ij}\). Also, there is an overall metric determinant factor \(\sqrt{\text{det}[(\beta^{-1} e^{\Phi})^2 h_{ij}]}\) . Note that all the quantities are normalized to be dimensionless. In our scaling limit, those background fields are the order of \footnote{\((\beta e^{-\Phi})^2 h^{ij}\) has non-zero components of order \(\mathcal{O}\Bigr(({\beta\over r})^{-2}\omega^4 \Bigr)\) for \(x^{i,j}\in \{\phi^1,\phi^2,\phi^3\}\). However, those components are contracted with \(\partial/\partial {\phi^{1,2,3}}\), and their contribution to the action vanishes since the background fields are independent of \(\phi^{1,2,3}\) coordinates. }
\begin{align}
&f_{ij}\sim\mathcal{O}\Bigr( ({\beta\over r})^0 \omega^{-1} \Bigr), \quad
F_{ij} \sim \Delta\cdot  \mathcal{O}\Bigr( ({\beta\over r})^0 \omega^{-1} \Bigr), \quad
\hat{R}^i{}_{jkl}\sim   \mathcal{O}\Bigr( ({\beta\over r})^0 \omega^{0} \Bigr), \quad
\beta A_0 \sim \Delta \cdot  \mathcal{O}\Bigr( ({\beta\over r})^0 \omega^0\Bigr), 
\nonumber \\
&\beta \Gamma^i{}_{j0}\sim   \mathcal{O}\Bigr( ({\beta\over r})^0 \omega^1 \Bigr), \quad
(\beta e^{-\Phi})^2 h^{ij}\sim \mathcal{O}\Bigr( ({\beta\over r})^{0} \omega^2 \Bigr), \quad
(\beta e^{-\Phi})^{-5} \sqrt{\text{det}[h_{ij}]}\sim \mathcal{O}\Bigr( ({\beta\over r})^1 \omega^{-6} \Bigr)
\label{eq304}
\end{align} 
Now let us consider a general form of the Lagrangian \(\mathcal{L}(n_1,n_2,n_3,n_4,n_5,n_6)\) which consists of \(n_1\) numbers of \(f_{ij}\), \(n_2\) numbers of \(F_{ij}\), \(n_3\) numbers of \(\hat{R}^i{}_{jkl}\), \(n_4\) numbers of \(\beta A_0\), \(n_5\) numbers of \(\beta \Gamma^i{}_{j0}\), and \(2n_6\) numbers of derivative \(\nabla_i\) where all \(n_i\)'s are non-negative integers. The indices are contracted with \(n_1+n_2+n_3+n_6\) numbers of \((\beta e^{-\Phi})^2 h^{ij}\). After integrating over \(S^5\), the lowest possible order of the action is
\begin{align}
\int_{S^5} \mathcal{L}(n_1,n_2,n_3,n_4,n_5,n_6)\sim 
\Delta^{n_2+n_4} \cdot \mathcal{O}\Bigr( ({\beta\over r})^1 \cdot \omega^{-6+n_1+n_2+2n_3+n_5+2n_6} \Bigr)
\label{eq309}
\end{align}
which is suppressed in our scaling limit due to the overall \((\beta/r)^1\) factor. Our result (\ref{eq309}) does not follow a naive derivative expansion argument that the \(n\)'th derivative term comes with an additional \(\beta^n\) power, since the background fields themselves consists \(\beta^{-1}\) factor. However, in our scaling limit, all non-CS terms are suppressed by the overall \(\beta\) factor, and higher derivative terms are more suppressed with higher \(\omega\) powers. Therefore, the thermal derivative expansion is valid for our background fields.
\par 
The result (\ref{eq309}) is based on the leading order counting of (\ref{eq304}). An honest evaluation of the action can give a smaller leading order if the cancellation happens at the expected leading order. In any case, the true leading order should be equal or subleading than the predicted leading order in (\ref{eq309}). For all non-CS terms on \(S^5\) evaluated in \cite{Choi:2018hmj}, we checked that they are consistent with our prediction.
\par 
Lastly, let us discuss \(W_\text{zero}\), which is the contribution from the dynamical zero-modes. The structure of the zero-modes is given by 5d SYM on the squashed \(S^5\). The 5d SYM has a coupling constant \(g^2_\text{YM}\sim \beta e^{-\Phi}\) which becomes \(g^2_\text{YM}\sim \mathcal{O}(r\omega)\) in our scaling limit. The 5d SYM is weakly coupled in the IR, and it is unlikely that the free energy has the terms proportional to the negative power of the coupling constant. Possibly the most severe IR divergence of the free energy would be \(\log g^2_\text{YM}\sim \mathcal{O}(\log \omega)\) \cite{Choi:2018hmj}. In this paper, we assume that \(W_\text{zero}\sim \mathcal{O}(\log \omega)\) in our scaling limit.
\par 
As a conclusion, we would like to emphasize that both \(W_\text{non-CS}\) and \(W_\text{zero}\) are suppressed in our scaling limit, but with different origins. \(W_\text{non-CS}\) is of order \(\mathcal{O}({\beta\over r})\), and it does not contribute to the index at the strict index point \({\beta/ r}\to 0\) given in (\ref{eq293_ind}). On the other hand, \(W_\text{zero}\) can contribute to the index, but it is subleading in the Cardy limit \(|\omega|\ll 1\). As a result, Cardy free energy can be obtained from the CS action as follows,
\begin{align}
\log I = -\lim_{\beta/r\to 0 } W_\text{CS}+\mathcal{O}(\log \omega).
\label{eq484_cardy}
\end{align}
For the anomaly polynomial given in (\ref{eq319_anom}), we obtain the following results,
\begin{align}
\log I=&-\Bigr({\mathfrak{a}\over 384}\Delta_R^4 +{\mathfrak{b} \pi^2\over 24} \Delta_R^2+{2\mathfrak{c} \pi^4\over 3} \Bigr){1\over \omega_1 \omega_2 \omega_3}
+\Bigr({\mathfrak{b}\over 96}\Delta_R^2 +{  (2\mathfrak{c}+\mathfrak{d}) \pi^2\over 6}\Bigr){\omega_1^2 +\omega_2^2+\omega_3^2\over \omega_1 \omega_2 \omega_3}
\nonumber \\
&+\mathcal{O}(\log \omega).
\label{eq470_cardy}
\end{align}
where the chemical potentials are constrained by \(\Delta_R-\omega_1 -\omega_2 -\omega_3=2\pi i\). The 't Hooft anomaly determines not only the leading order \(\mathcal{O}(\omega^{-3})\) term, but also the subleading corrections up to \(\mathcal{O}(\omega^{-1})\). The \(\mathcal{O}(\log \omega)\) correction includes the zero-mode contribution, which cannot be determined from the anomaly. In the following sections, we will compute the Cardy formula of various 6d SCFTs with (1,0) and (2,0) supersymmetries using the method explained so far.

\section{Cardy formulas of 6d SCFTs}
\label{section3}
As a preliminary, let us define the superconformal index and the modified index of 6d SCFTs and elaborate on the relation between them.
\par 
The superconformal index of 6d (2,0) SCFT is defined as follows,
\begin{align}
\mathcal{I}(\Delta_{1,2},\omega_{1,2,3})=\text{Tr}\Bigr[(-1)^F \cdot e^{-\beta'\{\mathcal{Q},\mathcal{Q}^\dagger\} } 
\cdot e^{-\Delta_1 Q_1-\Delta_2 Q_2} 
\cdot e^{-\omega_1 J_1-\omega_2 J_2-\omega_3 J_3}\Bigr].
\label{eq347_op}
\end{align}
The superconformal algebra of 6d (2,0) theories has a maximal bosonic subalgebra \(SO(6,2)\times SO(5)\) \cite{Nahm:1977tg}. The Cartans of \(SO(6,2)\) are the energy \(E\) and three angular momenta \(J_{1,2,3}\). The (2,0) R-symmetry is \(SO(5)\supset   SU(2)_L\times SU(2)_R\) whose Cartans are \(Q_{L,R}\). In (\ref{eq347_op}), \(Q_{1,2}\) are defined as \(Q_1={Q_R+Q_L}\) and \(Q_2={Q_R-Q_L}\). Sixteen Poincare supercharges of (2,0) theories are labelled by \(\mathcal{Q}^{s_1,s_2,s_3}_{t_1,t_2}\) where \(s_I,t_I=\pm 1\) and \(s_1 s_2 s_3 < 0\). The anticommutation relation between the supercharges is given as follows,
\begin{align}
\{ \mathcal{Q}^{s_1,s_2,s_3}_{t_1,t_2},(\mathcal{Q}^{s_1,s_2,s_3}_{t_1,t_2})^\dagger\}=E+s_1 J_1+s_2 J_2+s_3 J_3-2(t_1  Q_1+t_2 Q_2).
\end{align}
When the operators in the trace of (\ref{eq347_op}) are acted on the supercharge, we obtain the following factor,
\begin{align}
&e^{-\omega_1 J_1-\omega_2 J_2-\omega_3 J_3}\cdot e^{-\Delta_1 Q_1-\Delta_2 Q_2}  \cdot \mathcal{Q}^{s_1,s_2,s_3}_{t_1,t_2}
\nonumber \\
=& e^{{1\over 2}(s_1 \omega_1+s_2\omega_2+s_3\omega_3+t_1 \Delta_1 +t_2 \Delta_2) }\cdot \mathcal{Q}^{s_1,s_2,s_3}_{t_1,t_2} \cdot e^{-\omega_1 J_1-\omega_2 J_2-\omega_3 J_3}\cdot e^{-\Delta_1 Q_1-\Delta_2 Q_2}.
\end{align}
Let us define the superconformal index (\ref{eq347_op}) with the supercharge \(\mathcal{Q}\equiv \mathcal{Q}^{---}_{++}\). Then, the BPS states saturate the BPS bound given by
\begin{align}
E\geq J_1+J_2+J_3+4Q_R
\end{align}
and the chemical potentials are constrained by
\begin{align}
\Delta_R-\omega_1-\omega_2-\omega_3=0\quad (\text{mod }4\pi i)
\label{eq366_chemical}
\end{align}
where \(\Delta_R=\Delta_1+\Delta_2\) and \(\Delta_L=\Delta_1-\Delta_2\). As a result, the index is independent of the regulator \(\beta'\) and we can turn it off to zero.
\par 
Now, we introduce the modified version of the (2,0) index defined as follows,
\begin{align}
I(\Delta_{L,R},\omega_{1,2,3})=\text{Tr}\Bigr[ e^{-\omega_1 J_1-\omega_2 J_2-\omega_3 J_3}\cdot e^{-\Delta_R Q_R-\Delta_L Q_L} \Bigr]
\label{eq458_mod}
\end{align}
where the chemical potentials are constrained by \(\Delta_R-\omega_1-\omega_2-\omega_3=2\pi i\). Although the original index \(\mathcal{I}\) (\ref{eq347_op}) and the modified index \(I\) (\ref{eq458_mod}) are different in their expressions, they are equivalent up to the shift of the chemical potentials as follows,
\begin{align}
\mathcal{I}(\Delta_L ,\Delta_R,\omega_1,\omega_2,\omega_3)
=
I(\Delta_L,\Delta_R,\omega_1,\omega_2,\omega_3-2\pi i).
\label{eq477_shift}
\end{align}
The \(2\pi i\) shift of \(\omega_3\) in (\ref{eq458_mod}) yields \(e^{2\pi i J_3}\) factor. Since \(J_3\) is quantized to be integer for bosonic states and half-integer for fermionic states, one can replace \(e^{2\pi i J_3}\) by \((-1)^F\), thus proving (\ref{eq477_shift}). 
\par 
Now, let us move on to the indices of (1,0) SCFT. The superconformal index of 6d (1,0) SCFT is defined as follows,
\begin{align}
\mathcal{I}(\Delta,\omega_{1,2,3},x_i)=\text{Tr}\Bigr[(-1)^F \cdot e^{-\beta'\{\mathcal{Q},\mathcal{Q}^\dagger\}}\cdot e^{-\Delta_R  Q_R} \cdot e^{-\omega_1 J_1-\omega_2 J_2-\omega_3 J_3} \cdot \prod_i e^{-m_i F_i}\Bigr].
\label{eq713_index}
\end{align}
The superconformal algebra of 6d (1,0) theories has maximal bosonic subgroup \(SO(6,2)\times SU(2)_R\) \cite{Nahm:1977tg}. The Cartans of \(SO(6,2)\) are the energy \(E\) and three angular momenta \(J_{1,2,3}\). The Cartan of \(SU(2)_R\) R-symmetry is \(Q_R\). For (1,0) theories, there can be flavor symmetries with Cartan charges \(F_i\) and chemical potentials \(m_i\). Eight Poincare supercharges of (1,0) theories are labelled by \(\mathcal{Q}^{s_1,s_2,s_3}_t\) with \(s_I,t=\pm 1\) with \(s_1 s_2 s_3  < 0\). We take \(\mathcal{Q}\equiv \mathcal{Q}^{---}_+\) to define the index. Then the BPS bound is given by
\begin{align}
E\geq J_1+J_2+J_3+4Q_R
\end{align}
with the following chemical potential constraints,
\begin{align}
\Delta_R-\omega_1-\omega_2-\omega_3=0\quad (\text{mod }4\pi i).
\end{align}
Now, we introduce the modified version of the (1,0) index defined as follows,
\begin{align}
I=\text{Tr}\Bigr[  e^{-\Delta_R Q_R} \cdot e^{-\omega_1 J_1-\omega_2 J_2-\omega_3 J_3}\cdot \prod_i e^{-m_i F_i}\Bigr]
\label{eq725_modind}
\end{align}
where the chemical potentials are constrained by \(\Delta_R-\omega_1-\omega_2-\omega_3=2\pi i\). As well as the (2,0) index, the modified index and the original index of (1,0) SCFTs related as follows,
\begin{align}
\mathcal{I}(m_i,\Delta_R,\omega_1,\omega_2,\omega_3)
=
I(m_i,\Delta_R,\omega_1,\omega_2,\omega_3-2\pi i).
\end{align}

\subsection{(1,0) SCFT without flavor chemical potentials}
In this section, we consider a modified superconformal index of general 6d (1,0)  given as follows,
\begin{align}
I=\text{Tr}\Bigr[  e^{-\Delta_R Q_R} \cdot e^{-\omega_1 J_1-\omega_2 J_2-\omega_3 J_3}\Bigr]
\end{align}
where chemical potentials are constrained by \(\Delta_R-\omega_1-\omega_2-\omega_3=2\pi i\). The anomaly polynomial of the 6d SCFT takes the following form,
\begin{align}
P_8={1\over 4!}\Bigr(\mathfrak{a} \cdot c_2(R)^2+\mathfrak{b}\cdot c_2(R) p_1(T)+\mathfrak{c} \cdot p_1(T)^2+\mathfrak{d} \cdot p_2(T)  \Bigr)
\end{align}
with the theory dependent coefficients \(\mathfrak{a},\mathfrak{b},\mathfrak{c}\) and \(\mathfrak{d}\). The definitions of \(c_2(R)\) and \(p_{1,2}(T)\) are given in (\ref{eq329_c2}) and (\ref{eq335_p12}). From the above anomaly polynomial, we derived the following Cardy formula in section \ref{section22} as follows,
\begin{align}
\log I=&-\Bigr({\mathfrak{a}\over 384}\Delta_R^4 +{\mathfrak{b} \pi^2\over 24} \Delta_R^2+{2\mathfrak{c} \pi^4\over 3} \Bigr){1\over \omega_1 \omega_2 \omega_3}
+\Bigr({\mathfrak{b}\over 96}\Delta_R^2 +{  (2\mathfrak{c}+\mathfrak{d}) \pi^2\over 6}\Bigr){\omega_1^2 +\omega_2^2+\omega_3^2\over \omega_1 \omega_2 \omega_3}+\mathcal{O}(\log \omega)
\label{eq552_Cardy}
\end{align}
By inserting the chemical potential constraint \(\Delta_R=2\pi i+\omega_1+\omega_2+\omega_3\) , we can rearrange the Cardy formula (\ref{eq552_Cardy}) as follows,
\begin{align}
&\log I=-{\pi^4 (\mathfrak{a}-4\mathfrak{b}+16\mathfrak{c}) \over 24}{1\over \omega_1 \omega_2 \omega_3 }+{i\pi^3 (\mathfrak{a}-2\mathfrak{b})\over 12}{\omega_1+\omega_2+\omega_3\over \omega_1 \omega_2 \omega_3}
\nonumber \\
&+{\pi^2 (3\mathfrak{a}-4\mathfrak{b}+16\mathfrak{c}+8\mathfrak{d})\over 48}{\omega_1^2+\omega_2^2+\omega_3^2\over \omega_1 \omega_2 \omega_3}
+{\pi^2 (3\mathfrak{a}-2\mathfrak{b})\over 24}{\omega_1 \omega_2+\omega_2 \omega_3+\omega_3 \omega_1\over \omega_1 \omega_2 \omega_3 }
+\mathcal{O}(\log \omega)
\label{eq561_Cardy}
\end{align}
In this section, we will test our Cardy formula by comparing it with the known indices of 6d (1,0) free hyper and tensor multiplets. The indices are computed in the appendix \ref{appendix_B}.
\par 
The modified superconformal indices of (1,0) free hyper and tensor multiplets are given as follow,
\begin{align}
I_\text{hyp}&=\exp\sum_{n=1}^\infty{2\over n}{(-q_1 q_2 q_3)^n\over (1-q_1^{2n})(1-q_2^{2n})(1-q_3^{2n})}
\nonumber \\
I_\text{ten}&=\exp\sum_{n=1}^\infty{1\over n}{ (q_1 q_2 q_3)^{2n}-(q_1 q_2)^{2n}-(q_2 q_3)^{2n}-(q_3 q_1)^{2n} \over (1-q_1^{2n})(1-q_2^{2n})(1-q_3^{2n})}
\end{align}
where fugacities are defined as \(q_I=e^{-\omega_I/2}\). In the Cardy limit, the free energy can be expanded as follows,
\begin{align}
\log I_\text{hyp}&=\sum_{n=1}^\infty\Bigr[ {2(-1)^n\over n^4}{1\over \omega_1 \omega_2 \omega_3}-{(-1)^n\over 12n^2}{\omega_1^2+\omega_2^2+\omega_3^2\over \omega_1 \omega_2 \omega_3}+\mathcal{O}(\omega)\Bigr]
\nonumber \\
\log I_\text{ten}&=\sum_{n=1}^\infty \Bigr[ -{2\over n^4}{1\over \omega_1 \omega_2 \omega_3}-{1\over 6n^2}{\omega_1^2+\omega_2^2+\omega_3^2-3(\omega_1 \omega_2+\omega_2 \omega_3+\omega_3 \omega_1 ) \over \omega_1 \omega_2 \omega_3} -{1\over 2n}+\mathcal{O}(\omega)\Bigr].
\end{align}
As can be seen, the sum over \(n\) becomes divergent after \(\mathcal{O}(\omega^{-1})\) order. This divergence signals the breakdown of the perturbative expansion of small \(\omega\) after \(\mathcal{O}(\omega^{-1})\). We expect that after \(\mathcal{O}(\omega^{-1})\) order, there exist \(\mathcal{O}(\log \omega)\) terms that cannot be captured from the perturbative expansion. At least for the unrefined abelian (2,0) tensor index, the existence of those terms will be explicitly shown in the next subsection. Therefore, after summing over \(n\), we can expand our Cardy free energy as follows,
\begin{align}
\log I_\text{hyp}&=-{7\pi^4\over 360}{1\over \omega_1 \omega_2 \omega_3}+{\pi^2\over 144}{\omega_1^2+\omega_2^2+\omega_3^2\over \omega_1 \omega_2 \omega_3 }+\mathcal{O}(\log \omega)
\nonumber \\
\log I_\text{ten}&=-{\pi^4\over 45}{1\over \omega_1 \omega_2 \omega_3}-{\pi^2\over 36}{\omega_1^2+\omega_2^2+\omega_3^2\over \omega_1 \omega_2 \omega_3 }
+{\pi^2\over 12}{\omega_1 \omega_2+\omega_2 \omega_3+\omega_3 \omega_1\over \omega_1 \omega_2 \omega_3 }+\mathcal{O}(\log \omega).
\label{eq582_free}
\end{align}
Now, we compare the free energy (\ref{eq582_free}) with our Cardy formula (\ref{eq561_Cardy}). The anomaly coefficients of the free hyper and tensor multiplets are given by \cite{Ohmori:2014kda}
\begin{align}
\text{hyper: }& & &\mathfrak{a}=0, & &\mathfrak{b}=0, & &\mathfrak{c}={7\over 240}, & &\mathfrak{d}=-{1\over 60}
\nonumber \\
\text{tensor: }& & &\mathfrak{a}=1, & &\mathfrak{b}={1\over 2}, & &\mathfrak{c}={23\over 240}, & &\mathfrak{d}=-{29\over 60}
\label{eq589_anom}
\end{align}
By plugging the anomaly coefficients (\ref{eq589_anom}) to the Cardy formula (\ref{eq561_Cardy}), we can check that our Cardy formula gives the correct answers for all the result given in (\ref{eq582_free}).

\subsection{(2,0) SCFT of ADE type}
 6d (2,0) superconformal field theories have a well-known ADE classification \cite{Witten:1995zh,Seiberg:1996vs}. They are engineered from IIB string theory on \(\mathbb{C}^2/\Gamma\) where \(\Gamma\) is a discrete ADE subgroup of \(SU(2)\). For \(A_{N-1}\) SCFT with a single free tensor multiplet, it has the M-theory origin as a worldvolume theory of \(N\) coincident M5-branes \cite{Strominger:1995ac}. In this section, we study the superconformal indices of (2,0) theories in the Cardy limit using the background field method introduced in section \ref{section2}.
\par 
The background fields of 6d (2,0) SCFTs are the metric \(g_{\mu\nu}\) and the R-symmetry gauge field \(\mathbf{A}_{SU(2)}^{L,R}\). The explicit form of the metric and the graviphoton fields are written in (\ref{eq229_metric}) and (\ref{eq318_gravi}). We turn on Cartans of two \(SU(2)\) gauge fields \(\mathbf{A}_{SU(2)}^{L,R}\) of \(SU(2)_L \times SU(2)_R \subset SO(5)\) as follows,
\begin{align}
\mathbf{A}^{L,R}_{SU(2)}=\mathbf{A}^{L,R} \cdot T_{SU(2)}
\end{align}
where \(T_{SU(2)}\) is the Cartan generator of \(SU(2)\) normalized as \(\text{tr}_\textbf{adj}[(T_{SU(2)})^2]=2\). \(\mathbf{A}^{L,R}\) are \(U(1)\) Cartan gauge fields that have the following form,
\begin{align}
\mathbf{A}^{L,R}=A_0^{L,R} (d\tau+\mathbf{a})+\hat{\mathbf{A}}^{L,R} ,\quad A_0^{L,R}={\Delta_{L,R} \over \beta}.
\end{align}
At the end of the computation, we will insert \(\hat{\mathbf{A}}^{L,R} =-A_0^{L,R} \mathbf{a}\) to make the background gauge fields purely temporal.
\par 
As explained in section \ref{section2}, the Chern-Simons terms of the background effective action is determined from the anomaly polynomial. The anomaly polynomial of a 6d (2,0) SCFT of type \(G\)  is given by \cite{Ohmori:2014kda}
\begin{align}
P_8={h_G^\vee d_G\over 24} p_2(N)+{r_G\over 48}\Bigr[ p_2(N)-p_2(T)+{1\over 4}\Bigr(p_1(N)-p_1(T) \Bigr)^2 \Bigr].
\label{eq488_anom}
\end{align}
 Here, \(h_G^\vee\) is a dual Coxeter number, \(d_G\) is a group dimension, and \(r_G\) is a rank of \(G\). Various group-theoretic constants are listed in the table \ref{tab_group_const}.  The Pontryagin classes of the tangent bundle are defined in (\ref{eq335_p12}). For \(SO(5)\) normal bundle, its Pontryagin classes can be written in terms of Chern-classes of \(SU(2)_{L,R}\) as follows,
\begin{align}
&p_1(N)=-2\Bigr( c_2(L)+c_2(R) \Bigr),\quad 
p_2(N)=\Bigr(c_2(L)-c_2(R) \Bigr)^2 
\nonumber \\
&c_2(L,R)={1\over 4(2\pi)^2}\text{Tr}[(\mathbf{F}_{SU(2)}^{L,R})^2]
\label{eq408_normal}
\end{align}
where \(\mathbf{F}_{SU(2)}^{L,R}\) is a field strength two-form of \(\mathbf{A}_{SU(2)}^{L,R}\), and \(c_2\) is the second Chern-class. The capital trace is defined as \(\text{Tr}\equiv (h_G^\vee)^{-1} \text{tr}_\textbf{adj}\).
\par 
According to the replacement rule (\ref{eq273_repla}), the thermal anomaly polynomial \(P^T_8\) is given as follows,
\begin{align}
P_8^T={h_G^\vee d_G\over 24} p_2(N)+{r_G\over 48}\Bigr[ p_2(N)-\Bigr( p_2(T)-{\mathbf{F}_T^2\over 4\pi^2}p_1(T)\Bigr)+{1\over 4}\Bigr(p_1(N)-p_1(T)+{\mathbf{F}_T^2\over 4\pi^2} \Bigr)^2 \Bigr]
\label{eq555_ADEtherm}
\end{align}
where \(\mathbf{F}_T\) is the field strength two-form of the fictitious gauge field \(\mathbf{A}^T\) which was introduced in (\ref{eq234_fic}). The CS action on \(S^5\) are given as (\ref{eq229_cs}),
\begin{align}
W_\text{CS}=2\pi i \int_{\mathcal{N}^7} {dt+\mathbf{a}\over d\mathbf{a} }\Bigr[ P^T_8-\hat{P}^T_8 \Bigr]_{\mathbf{A}^T=0}
\end{align}
where \(\mathcal{N}^7=B^6\times S^1\) with a six-dimensional solid ball \(B^6\) with the boundary \(S^5\). After some computation, we obtain the following Chern-Simons action on \(S^5\),
\begin{align}
W_\text{CS}&={i\over \beta^3}\int_{S^5} \mathbf{a}d\mathbf{a}d\mathbf{a}\cdot \Bigr(- {h_G^\vee d_G+r_G\over 3072\pi^3}(\Delta_R^2-\Delta_L^2)^2
-{r_G\over 1536\pi^3} (\Delta_R^2+4\pi^2)(\Delta_L^2+4\pi^2)\Bigr)
\nonumber \\
&+{i\over \beta}\int_{S^5} \mathbf{a} \text{tr}[\hat{\mathbf{R}}^2]\cdot {r_G\over 3072\pi^3}(\Delta_R^2+\Delta_L^2-8\pi^2)+W_\text{grav-CS}.
\label{eq375_action}
\end{align}
The gravitational CS action \(W_\text{grav-CS}\) can be ignored in the Cardy limit. The integration over \(S^5\) can be evaluated by (\ref{eq329_CS}) and (\ref{eq367_mixed}). In the scaling limit (\ref{eq296_scaling}), we obtain
\begin{align}
&W_\text{CS}={h_G^\vee  d_G+r_G\over 384}{(\Delta_R^2-\Delta_L^2)^2\over \omega_1 \omega_2 \omega_3}
\nonumber \\
&+r_G\Bigr[{(\Delta_R^2 +4\pi^2)(\Delta_L^2+4\pi^2)\over 192\omega_1 \omega_2 \omega_3 }-{\Delta_R^2+\Delta_L^2-8\pi^2\over 192}{\omega_1^2 +\omega_2^2+\omega_3^2\over \omega_1 \omega_2 \omega_3} \Bigr]+W_\text{grav-CS}+\mathcal{O}({\beta\over r})
\label{eq331_csterms}
\end{align}
Then the Cardy free energy can be obtained from (\ref{eq331_csterms}) and (\ref{eq484_cardy}) as follows,
\begin{align}
&\log I=-{h_G^\vee  d_G+r_G\over 384}{(\Delta_R^2-\Delta_L^2)^2\over \omega_1 \omega_2 \omega_3}
\nonumber \\
&-r_G\Bigr[{(\Delta_R^2 +4\pi^2)( \Delta_L^2+4\pi^2)\over 192\omega_1 \omega_2 \omega_3}-{\Delta_R^2+\Delta_L^2-8\pi^2\over 192}{\omega_1^2+\omega_2^2+\omega_3^2\over \omega_1 \omega_2 \omega_3} \Bigr]+\mathcal{O}(\log \omega)
\label{eq410_cardy}
\end{align}
where the chemical potentials are constrained by \(\Delta_R-\omega_1-\omega_2-\omega_3=2\pi i\). 
\par 
The chemical potentials are periodic variables with \(4\pi i\) periodicity. In the Cardy limit, \(\Delta_L\) is the only relevant parameters with the periodicity since the other chemical potentials are fixed near \(\omega_{1,2,3}\simeq 0\) and \(\Delta_R\simeq 2\pi i\). We will see in the following paragraphs that for 6d (2,0) theories, our expression (\ref{eq410_cardy}) is valid in the following range which we shall call a `canonical chamber',
\begin{align}
-2\pi <\text{Im}[ \Delta_L]<2\pi 
\label{eq424_chamber}
\end{align}
The index outside the canonical chamber can be obtained from (\ref{eq410_cardy}) by \(4\pi i\) periodic shift of \(\Delta_L\).
\par 
Aside from the interacting 6d (2,0) SCFTs with ADE type, there is a free 6d (2,0) theory, which is made of the Abelian (2,0) tensor multiplet. For a single Abelian tensor multiplet, its anomaly polynomial is given as follows,
\begin{align}
P_8={1\over 48}\Bigr[ p_2(N)-p_2(T)+{1\over 4}\Bigr(p_1(N)-p_1(T) \Bigr)^2 \Bigr].
\label{eq547_free}
\end{align}
The anomaly polynomial (\ref{eq547_free}) takes the same form with (\ref{eq488_anom}) if we set \(h_G^\vee d_G=0\) and \(r_G=1\). Therefore, the Cardy free energy of the tensor multiplet is given as follows,
\begin{align}
&\log I=-{1\over 384}{(\Delta_R^2-\Delta_L^2)^2\over \omega_1 \omega_2 \omega_3}
\nonumber \\
&-\Bigr[{(\Delta_R^2 +4\pi^2)( \Delta_L^2+4\pi^2)\over 192\omega_1 \omega_2 \omega_3 }-{\Delta_R^2+\Delta_L^2-8\pi^2\over 192}{\omega_1^2+\omega_2^2+\omega_3^2\over \omega_1 \omega_2 \omega_3 } \Bigr]+\mathcal{O}(\log \omega).
\label{eq586_tensor}
\end{align}
\par 
The closed-form expression of the superconformal index of the free (2,0) tensor multiplet is known, and we can directly check our Cardy formula (\ref{eq586_tensor}). The superconformal index of the free tensor multiplet is known as follows,
\begin{align}
&\mathcal{I}_{U(1)}
=\text{Tr}\Bigr[ (-1)^F \cdot e^{-\Delta_L Q_L-\Delta_R Q_R}  \cdot e^{-\omega_1 J_1-\omega_2 J_2-\omega_3 J_3}\Bigr],\quad (\Delta_R-\omega_1 -\omega_2-\omega_3=0)
\nonumber \\
&=\exp\sum_{n=1}^\infty {1\over n}\Bigr[ {e^{-n{\Delta_R +\Delta_L\over 2} }+  e^{-n{\Delta_R -\Delta_L\over 2} }-e^{-n(\omega_1+\omega_2)}-e^{-n(\omega_2+\omega_3)}-e^{-n(\omega_3+\omega_1)}+e^{-n(\omega_1+\omega_2+\omega_3)} \over (1-e^{-n\omega_1})(1-e^{-n\omega_2})(1-e^{-n\omega_3}) } \Bigr].
\label{eq556_index}
\end{align}
As explained in (\ref{eq477_shift}), the superconformal index (\ref{eq556_index}) can be turned into the modified index by shifting the chemical potentials as \(\omega_3 \to \omega_3+2\pi i\). Then we obtain the following modified index,
\begin{align}
&I_{U(1)}
=\text{Tr}\Bigr[  e^{-\Delta_R Q_R-\Delta_L Q_L} \cdot e^{-\omega_1 J_1-\omega_2 J_2-\omega_3 J_3}\Bigr],\quad (\Delta_R -\omega_1 -\omega_2 -\omega_3=2\pi i)
\nonumber \\
&=\exp\sum_{n=1}^\infty {1\over n}\Bigr[ { e^{-n{\Delta_R +\Delta_L\over 2} }+  e^{-n{\Delta_R-\Delta_L\over 2} }-e^{-n(\omega_1+\omega_2)}-e^{-n(\omega_2+\omega_3)}-e^{-n(\omega_3+\omega_1)}+e^{-n(\omega_1+\omega_2+\omega_3)} \over (1-e^{-n\omega_1})(1-e^{-n\omega_2})(1-e^{-n\omega_3}) } \Bigr].
\label{eq465_index}
\end{align}
In the Cardy limit, after inserting \(\Delta_R=2\pi i +\omega_1+\omega_2+\omega_3\), the index be expanded as follows,
\begin{align}
&\log I_{U(1)}=\sum_{n=1}^\infty \Bigr[  { (-1)^n e^{n\Delta_L/2 }+(-1)^n e^{-n\Delta_L/2}-2\over n^4 \omega_1 \omega_2 \omega_3}
\nonumber \\
&+{-4\omega_1^2-4\omega_2^2-4\omega_2^2+12\omega_1 \omega_2+12\omega_2 \omega_3+12 \omega_3 \omega_1 -(-1)^n(e^{n\Delta_L/2}+e^{-n\Delta_L/2})(\omega_1^2+\omega_2^2+\omega_3^2 \over 24 n^2 \omega_1 \omega_2 \omega_3}
\Bigr)
\nonumber \\
&-{1\over 2n}+\mathcal{O}(\omega) \Bigr].
\label{eq593_naive}
\end{align}
The \(n\) summation becomes divergent after \(\mathcal{O}(\omega^{-1})\). This divergence signals the breakdown of the perturbative expansion of small \(\omega\) after \(\mathcal{O}(\omega^{-1})\). We expect that the free energy has \(\mathcal{O}(\log \omega)\) correction after \(\mathcal{O}(\omega^{-1})\) order which is consistent with our assumption of the zero-modes. The \(\mathcal{O}(\log \omega)\) correction can be explicitly seen in the index if we unrefine the chemical potentials as \(\omega_{1,2,3}=\omega\) and \(\Delta_L=2\pi i\). Then the modified index (\ref{eq465_index}) can be written as follows,
\begin{align}
I_\text{unrefined}=\exp\sum_{n=1}^\infty {1\over n}\Bigr[{2q^{3n}-3q^{4n}+q^{6n}\over (1-q^{2n})^3} \Bigr]=\prod_{k=3}^\infty (q^k;q^2)^{(-1)^k (k-1)},\quad q=e^{-\omega/2}
\label{eq597_asymp}
\end{align}
where the q-Pochhammer symbol is defined as \((a;q)=\prod_{k=0}^\infty (1-a q^k)\). The asymptotic expansion of the q-Pochhammer symbol for small \(\omega\) is \cite{Fredenhagen:2004cj}
\begin{align}
(q^k;q^2)=\exp\Bigr[-{\pi^2\over 6\omega}+{1-k\over 2}\log\omega +\mathcal{O}(\omega^0) \Bigr].
\label{eq643_asymp}
\end{align}
Using (\ref{eq643_asymp}), the infinite product over \(k\) in (\ref{eq597_asymp}) can be well-regulated, and we obtain the following asymptotic expansion of the unrefined free energy,
\begin{align}
\log I_\text{unrefine}=\exp\Bigr[{\pi^2 \over 8\omega }+{1\over 2}\log\omega+\mathcal{O}(\omega^0) \Bigr]
\end{align}
which is consistent with the perturbative expansion (\ref{eq593_naive}) up to \(\mathcal{O}(\omega^{-1})\). Therefore, we will trust our perturbative expansion (\ref{eq593_naive}) up to \(\mathcal{O}(\omega^{-1})\), keeping in mind the existence of \(\mathcal{O}(\log \omega)\) correction.
\par 
The infinite sum over \(n\) in (\ref{eq593_naive}) can be done with the polylogarithm functions as follows,
\begin{align}
&\log I_{U(1)}={1\over \omega_1\omega_2 \omega_3} \Bigr(\text{Li}_4(-e^{\Delta_L/2})+\text{Li}_4(-e^{-\Delta_L/2})-{\pi^4\over 45} \Bigr)
\nonumber \\
&+{2\pi^2 (-\omega_1^2-\omega_2^2-\omega_3^2+3\omega_1 \omega_2+3\omega_2 \omega_3+3\omega_3 \omega_1)-3(\omega_1^2+\omega_2^2+\omega_3^2)\Bigr( \text{Li}_2(-e^{\Delta_L/2})+\text{Li}_2(-e^{-\Delta_L/2})\Bigr)\over 72\omega_1 \omega_2 \omega_3}
\nonumber \\
&+\mathcal{O}(\log \omega).
\end{align}
We can further simplify the above expression by using the following polylogarithm identities,
\begin{align}
\text{Li}_m(-e^x) + (-1)^m \text{Li}_m(-e^{-x})=-{(2\pi i)^m\over m!} B_m\Bigr({1\over 2}+{x\over 2\pi i}\Bigr),\quad -\pi <\text{Im}[x]<\pi
\label{eq622_bernoulli}
\end{align}
where \(B_m\) is the \(m\)'th Bernoulli polynomial defined as \({t e^{xt}\over e^t-1}=\sum_{n=0}^\infty B_n (x) {t^n\over n!}\). In the canonical chamber \(-2\pi <\text{Im}[\Delta_L]<2\pi\) (\ref{eq424_chamber}), the index takes the following form,
\begin{align}
&\log I_{U(1)}=-{1\over 384}{(\Delta_L^2+4\pi^2)^2\over \omega_1 \omega_2 \omega_3}
\nonumber \\
&+{(\omega_1^2+\omega_2^2+\omega_3^2)\Delta_L^2 -4\pi^2 (\omega_1^2+\omega_2^2+\omega_3^2-4\omega_1 \omega_2 -4\omega_2 \omega_3-4\omega_3 \omega_1 )\over 192\omega_1 \omega_2 \omega_3 }+\mathcal{O}(\log \omega).
\label{eq459}
\end{align}
After inserting \(\Delta_R=2\pi i+\omega_1+\omega_2+\omega_3\) to the Cardy formula of the free tensor multiplet (\ref{eq586_tensor}), it is straightforward to check that (\ref{eq459}) and (\ref{eq586_tensor}) are exactly the same up to \(\mathcal{O}(\omega^{-1})\).

\subsection{E-string theory of arbitrary rank}
For (1,0) theories, there is no simple classification as ADE classification of (2,0) theories. The general (1,0) theories are engineered from F-theory on an elliptic Calabi-Yau threefold, and most of them admit atomic classification \cite{Heckman:2015bfa}. In this subsection and the next subsection, we will focus on two examples of (1,0) theories that can be engineered from M-theory as well.
\par 
A rank \(N\) E-string theory is the world volume theory of \(N\) M5-branes on a M9-brane \cite{Horava:1995qa,Klemm:1996hh}. The transverse space of M5-branes is \(\mathbb{R}^4\times \mathbb{R}_{>0}\) where \(\mathbb{R}^4\) is embedded in the M9-brane. The transverse \(\mathbb{R}^4\) directions host \(SO(4)\simeq SU(2)_R \times SU(2)_L\) global symmetry where \(SU(2)_R\) acts as the (1,0) R-symmetry and \(SU(2)_L\) acts as a flavor symmetry. Also, there is \(E_8\) global symmetry from the M9-brane. The superconformal index takes the following definition,
\begin{align}
I=\text{Tr}\Bigr[e^{-\omega_1 J_1-\omega_2 J_2-\omega_3 J_3}\cdot e^{-\Delta_R Q_R-\Delta_L Q_L} \cdot \prod_{a=1}^8 e^{- m_a F_a} \Bigr]
\end{align}
where the chemical potentials are constrained by \(\Delta_R-\omega_1-\omega_2-\omega_3=2\pi i\). Also, \(F_a\)'s are Cartans of \(E_8\) with chemical potentials \(m_a\).
\par 
The background gauge fields are \(\mathbf{A}_R\) for \(SU(2)_R\), \(\mathbf{A}_L\) for \(SU(2)_L\), and \(\mathbf{A}_{E}\) for \(E_8\) which take the following forms,
\begin{align}
\mathbf{A}_{L,R}={\Delta^{L,R}\over\beta} T_{SU(2)} d\tau,\quad 
\mathbf{A}_E=\sum_{a=1}^8 {m_a\over \beta} T^a_{E_8} d\tau
\end{align}
where \(T^a_{E_8}\)'s are Cartan generators of \(E_8\) normalized as \(\text{tr}_\textbf{adj}[ T^a_{E_8} T^b_{E_8}]=60 \delta_{ab}\). The anomaly polynomial of the rank \(N\) E-string theory is given as \cite{Ohmori:2014kda}
\begin{align}
P_8&={N^3-N\over 24}p_2(N)+{N\over 48}\Bigr[ p_2(N)-p_2(T)+{1\over 4}\Bigr( p_1(T)-p_1(N) \Bigr)^2 \Bigr]
+{N\over 2}\Bigr({N\over 2}\chi_4(N)+I_4 \Bigr)^2
\end{align}
where
\begin{align}
I_4={1\over 4}\Bigr({1\over (2\pi)^2}\text{Tr}[\mathbf{F}_E^2]+p_1(T)+p_1(N) \Bigr),\quad \chi_4(N)=c_2(L)-c_2(R)
\end{align}
and \(\mathbf{F}_E\) is a field strength two-form of \(\mathbf{A}_E\). The definitions for Pontryagin classes of the normal bundle and the tangent bundle can be found in (\ref{eq335_p12}) and (\ref{eq408_normal}). The thermal anomaly polynomial is given by
\begin{align}
P_8^T&={N^3-N\over 24}p_2(N)+{N\over 48}\Bigr[ p_2(N)-\Bigr(p_2(T)-{\mathbf{F}_T^2\over 4\pi^2}p_1(T)\Bigr)+{1\over 4}\Bigr( p_1(N)-p_1(T)+{\mathbf{F}_T^2\over 4\pi^2} \Bigr)^2 \Bigr]
\nonumber \\
&+{N\over 2}\Bigr[ {N\over 2}\chi_4(N)+{1\over 4}\Bigr(\text{Tr}[\mathbf{F}_E^2]+p_1(T)-{\mathbf{F}_T^2\over 4\pi^2}+p_1(N) \Bigr) \Bigr]^2.
\label{eq727_thermestring}
\end{align}
 Then the CS action of background fields on \(S^5\) is given as follows,
\begin{align}
W_\text{CS}=2\pi i \int_{\mathcal{N}^7} {d\tau+\mathbf{a}\over d\mathbf{a}}\Bigr[ P_8^T-\hat{P}_8^T\Bigr].
\end{align}
For the background fields introduced above, the CS action can be written as follows,
\begin{align}
&W_\text{CS}={i\over 256\pi^3 \beta^3 }\int_{S^5}\mathbf{a}d\mathbf{a}d\mathbf{a}\cdot \Bigr[ -{N^3\over 3} (\Delta_R^2-\Delta_L^2)^2
+{N^2\over 2} (\Delta_R^2-\Delta_L^2)(4\sum_a m_a^2+8\pi^2-\Delta_R^2-\Delta_L^2)
\nonumber \\
&-{N\over 12}\Bigr(3(\Delta_R^4+\Delta_R^4)+8\Delta_R^2 \Delta_L^2-40\pi^2 (\Delta_R^2+\Delta_L^2) +24(8\pi^2-\Delta_R^2-\Delta_L^2)\sum_a m_a^2+48 (\sum_a m_a^2)^2+224\pi^4 \Bigr) \Bigr]
\nonumber \\
&+{i\over3072\pi^3 \beta}\int_{S^5} \mathbf{a}\text{tr}[\hat{\mathbf{R}}^2]\cdot 
\Bigr[-6N^2 (\Delta_R^2-\Delta_L^2)+N\Bigr(40\pi^2 -5(\Delta_R^2+\Delta_L^2)+24\sum_a m_a^2 \Bigr)  \Big]
\nonumber \\
&+W_\text{grav-CS}.
\end{align}
Again, pure gravitational CS terms can be ignored in the Cardy limit. After evaluating the integral over \(S^5\), the Cardy free energy can be obtained as follows,
\begin{align}
&\log I={1\over 32\omega_1 \omega_2 \omega_3}\Bigr[ -{N^3\over 3} (\Delta_R^2-\Delta_L^2)^2
+{N^2\over 2} (\Delta_R^2-\Delta_L^2)(4\sum_a m_a^2+8\pi^2-\Delta_R^2-\Delta_L^2)
\nonumber \\
&-{N\over 12}\Bigr(3(\Delta_R^4+\Delta_L^4)+8\Delta_R^2 \Delta_L^2-40\pi^2 (\Delta_R^2+\Delta_L^2) +24(8\pi^2-\Delta_R^2-\Delta_L^2)\sum_a m_a^2+48 (\sum_a m_a^2)^2+224\pi^4 \Bigr) \Bigr]
\nonumber \\
&+{\omega_1^2+\omega_2^2+\omega_3^2 \over 192\omega_1 \omega_2 \omega_3}
\Bigr[-6N^2 (\Delta_R^2-\Delta_L^2)+N\Bigr(40\pi^2 -5(\Delta_R^2+\Delta_L^2)+24\sum_a m_a^2 \Bigr)  \Big]
+\mathcal{O}(\log \omega)
\label{eq766_estring}
\end{align}
where chemical potentials are constrained by \(\Delta_R-\omega_1-\omega_2-\omega_3=2\pi i\).

\subsection{M5 branes on ALE singularities}
In this section, we consider another (1,0) SCFT, which is the worldvolume theory of \(N\) M5-branes probing \(\mathbb{C}^2/\Gamma_G\) singularity. It  has the following quiver theory description,
\begin{align}
\boxed{G_0} - G_1 - ... - G_{N-1} - \boxed{G_N}
\end{align}
where \(G_{1,...,N-1}\) are gauge symmetries and \(G_{0,N}\) are flavor symmetries of the same simply laced Lie group \(G\). The line \(-\) denotes a `conformal matter' which is a weakly coupled hyper multiplet if \(G=A_N\), but becomes another nontrivial 6d SCFT with a fractionalized M5-brane when \(G\) is \(D_N\) or \(E_N\) \cite{DelZotto:2014hpa}. We also have vector multiplets for each of the gauge node and a tensor multiplet as a center of the mass degrees of freedom. The 6d theory has \(SU(2)\) R-symmetry and \(G_0\times G_N\) flavor symmetry. The modified superconformal index is defined as follows,
\begin{align}
I=\text{Tr}\Bigr[e^{-\Delta_R Q_R }\cdot e^{-\omega_ 1J_1 -\omega_2 J_2-\omega_3 J_3} \cdot \prod_{i=1}^{r_G} e^{-x_i M_i-y_i N_i}\Bigr]
\label{eq485_index}
\end{align}
where chemical potentials are constrained by \(\Delta_R-\omega_1-\omega_2-\omega_3=2\pi i\). Here, \(M_i\)'s are Cartans of \(G_0\) and \(N_i\)'s are Cartans of \(G_N\). \(x_i\)'s and \(y_i\)'s are their chemical potentials respectively.
\par 
The background gauge fields \(\mathbf{A}_R\) for \(SU(2)_R\) R-symmetry and \(\mathbf{A}_{0,N}\) for \(G_0\) and \(G_N\) flavor symmetry are turned on as
\begin{align}
\mathbf{A}_R={\Delta_R\over \beta}T_{SU(2)} d\tau,\quad
\mathbf{A}_{0}&={x_i\over\beta} T_G^i d\tau ,\quad 
\mathbf{A}_{N}={y_i\over\beta} T_G^i d\tau 
\end{align}
where \(T_G^i\)'s are the Cartan generators of \(G\). The anomaly polynomial of this theory is given by \cite{Ohmori:2014kda}
\begin{align}
P_8&={|\Gamma|^2 N^3\over 24}c_2(R)^2-{N\over 48}c_2(R)\Bigr( |\Gamma|(r_G+1)-1\Bigr) \Bigr(4c_2(R)+p_1(T)\Bigr)
\nonumber \\
&-{N\over 8 (2\pi)^2  }|\Gamma| c_2(R)\Bigr(\text{Tr}[\mathbf{F}_0^2]+\text{Tr}[\mathbf{F}_N^2] \Bigr)
+{N\over 8}\Bigr({1\over 6}c_2(R) p_1(T)-{1\over 6}p_2(T)+{1\over 24}p_1(T)^2 \Bigr)
\nonumber \\
&-{1\over 2}P_8^\text{vec}(\mathbf{F}_0)-{1\over 2}P_8^\text{vec}(\mathbf{F}_N)
-{1\over 2(2\pi)^4  N}\Bigr({1\over 4}\text{Tr}[\mathbf{F}_0^2]-{1\over 4}\text{Tr}[\mathbf{F}_N^2] \Bigr)^2
\end{align}
where \(\mathbf{F}_{0,N}\) is a field strength of the flavor symmetry \(G_{0,N}\). The anomaly polynomial of a vector multiplet \(P_8^\text{vec}\) is
\begin{align}
P_8^\text{vec}(\mathbf{F})=&-{1\over 24}\Bigr( {h_G^\vee\over (2\pi)^4} \text{Tr}[\mathbf{F}^4]+{6 h_G^\vee\over (2\pi)^2} c_2(R) \text{Tr}[\mathbf{F}^2]+d_G c_2(R)^2 \Bigr)
\nonumber \\
&-{1\over 48}\Bigr({h_G^\vee\over (2\pi)^2} \text{Tr}[\mathbf{F}^2]+d_G c_2(R)\Bigr) p_1(T)
-d_G {7p_1(T)^2-4p_2(T)\over 5760}.
\end{align}
The various group theoretic parameters are listed in the table \ref{tab_group_const}.
\begin{table}[t!]
\centering
\begin{tabular}{|c||c|c|c|c|c|} \hline
 & \(SU(k)\) & \(SO(2k)\) & \(E_6\) & \(E_7\) & \(E_8\)  \\ \hline \hline
\(r_G\) & \(k-1\) & \(k\) & 6 & 7 & 8 \\ \hline
\(h_G^\vee\) & \(k\) & \(2k-2\)  &12 & 18 & 30 \\ \hline
\(d_G\) & \(k^2-1\) & \(k(2k-1)\) & 78 & 133 & 248 \\ \hline
\(s_G\) & \(1/2\) & \(1\)  & 3 & 6 & 30 \\ \hline
\(t_G\) & \(2k\) & \(2k-8\) & 0 & 0 & 0  \\ \hline
\(u_G\) & 2 & 4 & 6&  8 & 12 \\ \hline
\(|\Gamma|\) & \(k\) & \(4k-8\) & 24 & 48 & 120 \\ \hline
\end{tabular}
\caption{Group-theoretic constants for ADE type. We follow the convention of \cite{Ohmori:2014kda}.}
\label{tab_group_const}
\end{table}
\par 
Now it is straight forward to compute the CS action of the background fields as follows,
\begin{align}
W_\text{CS}=2\pi i \int_{\mathcal{N}^7} {dt+\mathbf{a}\over d\mathbf{a} }\Bigr[ P^T_8-\hat{P}^T_8 \Bigr]_{\mathbf{A}^T=0}.
\end{align}
For the background fields introduced above, the CS action is given as follows,
\begin{align}
&W_\text{CS}=-{i\over 3072\pi^3 \beta^3}\int_{S^5}\mathbf{a}d\mathbf{a}d\mathbf{a}\cdot \Biggr( N^3 |\Gamma|^2 \Delta_R^4 
\nonumber \\
&-N\Bigr[ 2\Bigr( |\Gamma|(r_G+1)-1\Bigr)\Delta_R^4
+4\Delta_R^2 \Bigr(2\pi^2(|\Gamma|(r_G+1)-2)+3|\Gamma| (T_2(x)+T_2(y))  \Bigr)-32\pi^4 \Bigr]
\nonumber \\
&+ d_G \Delta_R^4+ \Delta_R^2 \Bigr( 8\pi^2 d_G+12 h_G (T_2(x)+T_2(y)) \Bigr)+8h_G \Bigr(2\pi^2 (T_2(x)+T_2(y))+T_4(x)+T_4(y) \Bigr) 
\nonumber \\
&+{112\pi^4\over 15}d_G  -{12\over N}\Bigr(T_2(x)-T_2(y)\Bigr)^2
\Biggr)
\nonumber \\
&-{i\over 3072 \pi^3 \beta}\int_{S^5}\mathbf{a}\text{tr}[\hat{\mathbf{R}}^2]\cdot \Bigr[
N\Bigr( |\Gamma| (r_G+1)-2 \Bigr) \Delta_R^2+8\pi^2 \Bigr)
-\Bigr( d_G \Delta_R^2+2h_G (T_2(x)+T_2(y))+{4\pi^2\over 3}d_G \Bigr)
\Bigr]
\nonumber \\
&+W_\text{grav-CS}
\end{align}
where the function \(T_n\) is defined as
\begin{align}
T_n(x)=\text{Tr}\Bigr[\sum_{i=1}^{r_G}( x_i T_G^i)^n \Bigr].
\end{align}
Again, the pure gravitational CS terms are ignored in the Cardy limit. Then, we obtain the following expression for the Cardy free energy,
\begin{align}
&\log I =-{1\over 384\omega_1 \omega_2 \omega_3}\Biggr( N^3 |\Gamma|^2 \Delta_R^4 
\nonumber \\
&-N\Bigr[ 2\Bigr( |\Gamma|(r_G+1)-1\Bigr)\Delta_R^4
+4\Delta_R^2 \Bigr(2\pi^2(|\Gamma|(r_G+1)-2)+3|\Gamma|(T_2(x)+T_2(y)) \Bigr)-32\pi^4 \Bigr]
\nonumber \\
&+ d_G \Delta_R^4+ \Delta_R^2 \Bigr( 8\pi^2 d_G+12 h_G (T_2(x)+T_2(y))\Bigr)+8h_G \Bigr(2\pi^2 (T_2(x)+T_2(y))+T_4(x)+T_4(y) \Bigr) 
\nonumber \\
&+{112\pi^4\over 15}d_G -{12\over N}\Bigr(T_2(x)-T_2(y)\Bigr)^2
\Biggr)
\nonumber \\
&-{\omega_1^2+\omega_2^2+\omega_3^2\over 192\omega_1 \omega_2 \omega_3}\Bigr[
N\Bigr( (|\Gamma| (r_G+1)-2 ) \Delta_R^2+8\pi^2 \Bigr)
-\Bigr( d_G \Delta_R^2+2h_G(T_2(x)+T_2(y))+{4\pi^2\over 3}d_G \Bigr)
\Bigr]
\nonumber \\
&+\mathcal{O}(\log \omega)
   \label{eq568_results}
\end{align}
where the chemical potentials constrained by \(\Delta_R-\omega_1-\omega_2-\omega_3=2\pi i\).
\par 
Now, in order to check our Cardy formula (\ref{eq568_results}), let us consider a single M5-brane probing \(\mathbb{C}^2/\mathbb{Z}_k\) singularity which has the following quiver description,
\begin{align}
\boxed{SU(k)_0} - \boxed{SU(k)_1}.
\end{align}
The above quiver theory is described by a single free (1,0) tensor multiplet and a free hyper multiplet charged under the two flavor symmetries \(SU(k)_{0,1}\). The superconformal index is given by \cite{Benvenuti:2016dcs}
\begin{align}
\mathcal{I}&=\text{Tr}\Bigr[(-1)^F\cdot e^{-\omega_1 J_1-\omega_2 J_2-\omega_3 J_3)}\cdot e^{-\Delta_R Q_R }\cdot \prod_{i=1}^{k-1} e^{-x_i M_i-y_i N_i} \Bigr],\quad \Bigr(\Delta_R -\omega_1 -\omega_2 -\omega_3 =0\Bigr)
\nonumber \\
&=\exp \sum_{n=1}^\infty {1\over n}{1\over (1-e^{-n\omega_1}) (1-e^{-n\omega_2}) (1-e^{-n\omega_3})}
\Bigr( e^{-n(\omega_1+\omega_2+\omega_3)}-e^{-n(\omega_1+\omega_2) }-e^{-n(\omega_2+\omega_3) }-e^{-n(\omega_3+\omega_1) }
\nonumber \\
&+e^{-n{\Delta_R \over 2} }  \chi(nx) \cdot \bar{\chi}(ny)+e^{-n{\Delta_R \over 2} } \bar{\chi}(nx) \cdot  \chi(ny)\Bigr). 
\label{eq581_index}
\end{align}
Here \(Q_R\) is the \(SU(2)\) R-symmetry Cartan. \(M_i\)'s and \(N_i\)'s are \(SU(k)_0\) and \(SU(k)_1\) charges respectively. For \(k=1\), the index (\ref{eq581_index}) reduces to the index of the free (2,0) tensor multiplet (\ref{eq556_index}). \(\chi\) is the fundamental character of \(SU(k)\), and \(\bar{\chi}\) is the anti-fundamental character. It is convenient to define a new set of basis \(X^i\)'s for the \(SU(k)_0\) chemical potentials as follows,
\begin{align}
x_i T^i_\textbf{fund}&=\text{diag}\Bigr( X_1,X_2,...,X_k \Bigr)
\end{align}
where \(X_i\)'s satisfy \(\sum_{i=1}^k X_k=0\). We define \(Y_i\)'s in a similar way as \(X_i\)'s. Then, the (anti-)fundamental characters are written as
\begin{align}
\chi(nx)=\sum_{i=1}^k e^{-n X_k},\quad \bar{\chi}(nx)=\sum_{i=1}^k e^{n X_k}.
\end{align}
The index (\ref{eq581_index}) can be turned into the modified index by shifting \(\omega_3 \to \omega_3+2\pi i\). Then, the modified index is given as follows,
\begin{align}
& I=\text{Tr}\Bigr[e^{-\omega_1 J_1-\omega_2 J_2-\omega_3 J_3}\cdot e^{-\Delta_R Q_R}\cdot \prod_{i=1}^{k-1} e^{-x_i M_i-y_i N_i} \Bigr],\quad 
\Bigr( \Delta_R -\omega_1-\omega_2-\omega_3=2\pi i\Bigr)
\nonumber \\
&=\exp \sum_{n=1}^\infty {1\over n}{1\over (1-e^{-n\omega_1}) (1-e^{-n\omega_2}) (1-e^{-n\omega_3})}
\Bigr( e^{-n(\omega_1+\omega_2+\omega_3)}-e^{-n(\omega_1+\omega_2) }-e^{-n(\omega_2+\omega_3) }-e^{-n(\omega_3+\omega_1) }
\nonumber \\
&+e^{-n{\Delta_R \over 2} }  \chi[nx] \cdot \bar{\chi}[ny]+e^{-n{\Delta_R \over 2} } \bar{\chi}[nx] \cdot  \chi[ny]\Bigr) .
\end{align}
The small \(\omega\) expansion of the free energy is
\begin{align}
&\log I=\sum_{n=1}^\infty \Bigr[ {-2+   (-1)^n \Bigr( \chi[nx] \cdot \bar{\chi}[ny]+\bar{\chi}[nx] \cdot  \chi[ny] \Bigr) \over n^4 \omega_1 \omega_2 \omega_3 }
\nonumber \\
&+{1\over  24n^2 \omega_1 \omega_2 \omega_3}\Bigr( -4\omega_1^2-4\omega_2^2-4\omega_3^2+12\omega_1 \omega_2+12 \omega_2 \omega_3+12 \omega_3 
\nonumber \\
&-(-1)^n (\omega_1^2+\omega_2^2+\omega_3^2) (   \chi[nx] \cdot \bar{\chi}[ny]+  \bar{\chi}[nx] \cdot  \chi[ny] )  \Bigr)
-{1\over 2n}+\mathcal{O}(\omega)\Bigr].
\end{align}
Again, we encounter a divergent term which signals the breakdown of the perturbative expansion after \(\mathcal{O}(\omega^{-1})\) order. As we have seen, the subleading terms after \(\mathcal{O}(\omega^{-1})\) contain \(\mathcal{O}(\log \omega)\) correction. However, the terms up to \(\mathcal{O}(\omega^{-1})\) is sufficient to check the validity of (\ref{eq568_results}). The infinite \(n\) summations can be done by using the polylogarithm functions as follows,
\begin{align}
\log I&={1\over \omega_1 \omega_2 \omega_3}\Bigr[ -{\pi^4\over 45}+\sum_{i,j=1}^k \Bigr(\text{Li}_4(-e^{-X_i+Y_j}) +\text{Li}_4(-e^{X_i-Y_j}) \Bigr) \Bigr]
\nonumber \\
&+{1\over 24\omega_1 \omega_2 \omega_3}\Bigr[ {\pi^2\over 6}\Bigr(-4\omega_1^2-4\omega_2^2-4\omega_3^2+12\omega_1 \omega_2 +12 \omega_2 \omega_3 +12 \omega_3 \omega_1 \Bigr) 
\nonumber \\
&-(\omega_1^2+\omega_2^2+\omega_3^2) \sum_{i,j=1}^k \Bigr(\text{Li}_2(-e^{-X_i+Y_j}) +\text{Li}_2(-e^{X_i-Y_j}) \Bigr)\Bigr]+\mathcal{O}(\log \omega).
\end{align}
It can be further simplified by using the polylogarithm identity (\ref{eq622_bernoulli}). In the canonical chamber where \(-\pi<\text{Im}[X_i-Y_j]<\pi\), the Cardy free energy can be written as follows,
\begin{align}
\log I&={1\over \omega_1 \omega_2 \omega_3}\Bigr[ -{\pi^4\over 45}-{(2\pi i)^4\over 4!}\sum_{i,j=1}^k B_4 \Bigr({1\over 2}+{X_i-Y_j\over 2\pi i} \Bigr) \Bigr]
\nonumber \\
&+{1\over 24\omega_1 \omega_2 \omega_3}\Bigr[ {\pi^2\over 6}\Bigr(-4\omega_1^2-4\omega_2^2-4\omega_3^2+12\omega_1 \omega_2 +12 \omega_2 \omega_3 +12 \omega_3 \omega_1 \Bigr)
\nonumber \\
&+(\omega_1^2+\omega_2^2+\omega_3^2)\cdot {(2\pi i)^2\over 2!}  \sum_{i,j=1}^k B_2\Bigr({1\over 2}+{X_i-Y_j\over 2\pi i} \Bigr)\Bigr]+\mathcal{O}(\log\omega)
\label{eq891_ddd}
\end{align}
Now, let us compare (\ref{eq891_ddd}) with our result (\ref{eq568_results}). For \(G=SU(k)\),  \(T_n(x)\) in (\ref{eq568_results}) can be written as
\begin{align}
T_2(x)&={1\over s_G}\text{tr}_\textbf{fund}[(x_i T^i)^2]=2\sum_{i=1}^k X_i^2
\nonumber \\
T_4(x)&={t_G\over h_G^\vee}\text{tr}_\textbf{fund}[(x_i T^i)^4]+{3u_G\over 4h_G^\vee s_G^2} \text{tr}_\textbf{fund}[(x_i T^i)^2]^2
=2\sum_{i=1}^k X_i^4+{6\over k}(\sum_{i=1}^k X_i^2)^2
\end{align}
where we used the trace identities in \cite{Ohmori:2014kda}. Then we can check that (\ref{eq891_ddd}) and (\ref{eq568_results}) are exactly the same up to \(\mathcal{O}(\omega^{-1})\) order.

\section{Equivariant integral of the thermal anomaly polynomial}
\label{section4}
In \cite{Bobev:2015kza}, it was conjectured that the supersymmetric Casimir energy can be obtained from the equivariant integral of the anomaly polynomial. In this section, we apply their conjecture to the Cardy formula such that it is given by the equivariant integral of the thermal anomaly polynomial. As a result, we find that the Cardy formula and the equivariant integral of the thermal anomaly polynomial are related as
\begin{align}
\log I=-\int P_8^T + \mathcal{O}(\log \omega)
\label{eq925_conjecture}
\end{align}
where \(\int\) denotes the equivariant integral. Note that the Cardy limit \(|\omega_i|\ll 1\) is assumed.
\par 
The equivariant integral is defined from the equivariant cohomology, where the de Rham differential is twisted with the equivariant parameters. The anomaly polynomial \(P_{d+2}\) can be viewed as an equivariant form on \(\mathbb{R}^d\) with equivariant parameters given by the chemical potentials of the Cartan subalgebra of the global symmetries. See \cite{Bobev:2015kza} for the detailed explanation. The equivariant integral of \(P_{d+2}\) is evaluated by the Duistermaat-Heckman formula \cite{Duistermaat:1982vw} given as follows,
\begin{align}
\int P_{d+2}=\sum_{p\in \text{fixed points}} {P_{d+2}\over e(TM)}\Bigr|_{p}
\label{eq1013_DH}
\end{align}
where \(e(TM)\) is the equivariant Euler class. In our case, there is only a single fixed point \(p_0\), and the 0-form components of the equivariant forms at the fixed point are given as follows \cite{Bobev:2015kza},
\begin{align}
e(TM)|_{p_0}&=\omega_1 \omega_2 \omega_3, &
c_2(R)|_{p_0}&=-{\Delta_R^2\over 4}
\nonumber \\
p_1(T)|_{p_0}&= \omega_1^2+\omega_2^2+\omega_3^2, &
p_2(T)|_{p_0}&=\omega_1^2 \omega_2^2 +\omega_2^2 \omega_3^2+\omega_3^2\omega_1^2
\label{eq928_rule1}
\end{align}
For the thermal anomaly polynomial, there is the fictitious gauge field \(\mathbf{A}_T\). The Chern class of the fictitious gauge field at the fixed point is given as follows,
\begin{align}
{1\over 4\pi^2}\mathbf{F}_T^2\Bigr|_{p_0}= -(2\pi i)^2.
\label{eq938_rule3}
\end{align}
Then, the equivariant integral of the thermal anomaly polynomial (\ref{eq319_anom}) is given by
\begin{align}
-\int P_8^T=&-\Bigr({\mathfrak{a}\over 384}\Delta_R^4 +{\mathfrak{b} \pi^2\over 24} \Delta_R^2+{2\mathfrak{c} \pi^4\over 3} \Bigr){1\over \omega_1 \omega_2 \omega_3}
+\Bigr({\mathfrak{b}\over 96}\Delta_R^2 +{\pi^2  (2\mathfrak{c}+\mathfrak{d}) \over 6}\Bigr){\omega_1^2 +\omega_2^2+\omega_3^2\over \omega_1 \omega_2 \omega_3}
\nonumber \\
&-{\mathfrak{d}\over 24} {\omega_1^2 \omega_2^2 + \omega_2^2 \omega_3^2+\omega_3^2 \omega_1^2 \over \omega_1 \omega_2 \omega_3}
\label{eq1034_equiv}
\end{align}
Comparing (\ref{eq1034_equiv}) with the Cardy formula (\ref{eq470_cardy}), we can check that our relation (\ref{eq925_conjecture}) is valid. Note that \(\mathcal{O}(\omega^1)\) term in the equivariant integral can be ignored when computing the Cardy free energy since it is subleading than \(\mathcal{O}(\log \omega)\) correction in (\ref{eq925_conjecture}).
\par 
Now, let us consider 6d (2,0) SCFTs with ADE type. The equivariant integral of the thermal anomaly polynomial of 6d (2,0) SCFT (\ref{eq555_ADEtherm}) is given by,
\begin{align}
-\int P_8^T=&-{h_G^\vee  d_G+r_G\over 384}{(\Delta_R^2-\Delta_L^2)^2\over \omega_1 \omega_2 \omega_3}
-r_G\Bigr[{(\Delta_R^2 +4\pi^2)( \Delta_L^2+4\pi^2)\over 192\omega_1 \omega_2 \omega_3}
\nonumber \\
&-{\omega_1^2+\omega_2^2+\omega_3^2\over \omega_1 \omega_2 \omega_3} {\Delta_R^2+\Delta_L^2-8\pi^2\over 192}
+{\omega_1^4+\omega_2^4+\omega_3^4-2\omega_1^2 \omega_2^2-2\omega_2^2 \omega_3^2-2\omega_3^2 \omega_1^2\over 192\omega_1 \omega_2 \omega_3}\Bigr].
\label{eq952_equiv}
\end{align}
Comparing (\ref{eq952_equiv}) with the Cardy formula (\ref{eq410_cardy}), we can check that our relation (\ref{eq925_conjecture}) is valid for general (2,0) theories.
\par 
Now, we move on to the 6d (1,0) theories. For the rank \(N\) E-string theory, the equivariant integral of the thermal anomaly polynomial is given by,
\begin{align}
&-\int P_8^T={1\over 32\omega_1 \omega_2 \omega_3}\Bigr[ -{N^3\over 3} (\Delta_R^2-\Delta_L^2)^2
+{N^2\over 2} (\Delta_R^2-\Delta_L^2)(4\sum_a m_a^2+8\pi^2-\Delta_R^2-\Delta_L^2)
\nonumber \\
&-{N\over 12}\Bigr(3(\Delta_R^4+\Delta_L^4)+8\Delta_R^2 \Delta_L^2-40\pi^2 (\Delta_R^2+\Delta_L^2) +24(8\pi^2-\Delta_R^2-\Delta_L^2)\sum_a m_a^2+48 (\sum_a m_a^2)^2+224\pi^4 \Bigr) \Bigr]
\nonumber \\
&+{\omega_1^2+\omega_2^2+\omega_3^2 \over 192\omega_1 \omega_2 \omega_3}
\Bigr[-6N^2 (\Delta_R^2-\Delta_L^2)+N\Bigr(40\pi^2 -5(\Delta_R^2+\Delta_L^2)+24\sum_a m_a^2 \Bigr)  \Big]
\nonumber \\
&-N{7(\omega_1^2+\omega_2^2+\omega_3^2)+10(\omega_1^2 \omega^2+\omega_2^2\omega_3^2+\omega_3^2\omega_1^2)\over 192\omega_1 \omega_2 \omega_3}.
\end{align}
For M5-branes on ALE singularities, the equivariant integral of the thermal anomaly polynomial is given by,
\begin{align}
&-\int P_8^T =-{1\over 384\omega_1 \omega_2 \omega_3}\Biggr( N^3 |\Gamma|^2 \Delta_R^4 
\nonumber \\
&-N\Bigr[ 2\Bigr( |\Gamma|(r_G+1)-1\Bigr)\Delta_R^4
+4\Delta_R^2 \Bigr(2\pi^2(|\Gamma|(r_G+1)-2)+3|\Gamma|(T_2(x)+T_2(y)) \Bigr)-32\pi^4 \Bigr]
\nonumber \\
&+ d_G \Delta_R^4+ \Delta_R^2 \Bigr( 8\pi^2 d_G+12 h_G (T_2(x)+T_2(y))\Bigr)+8h_G \Bigr(2\pi^2 (T_2(x)+T_2(y))+T_4(x)+T_4(y) \Bigr) 
\nonumber \\
&+{112\pi^4\over 15}d_G -{12\over N}\Bigr(T_2(x)-T_2(y)\Bigr)^2
\Biggr)
\nonumber \\
&-{\omega_1^2+\omega_2^2+\omega_3^2\over 192\omega_1 \omega_2 \omega_3}\Bigr[
N\Bigr( (|\Gamma| (r_G+1)-2 ) \Delta_R^2+8\pi^2 \Bigr)
-\Bigr( d_G \Delta_R^2+2h_G(T_2(x)+T_2(y))+{4\pi^2\over 3}d_G \Bigr)
\Bigr]
\nonumber \\
&+{N\over 192}{2\omega_1^2 \omega_2^2+2\omega_2^2 \omega_3^2+2\omega_3^2 \omega_1^2-\omega_1^4-\omega_2^4-\omega_3^4\over \omega_1 \omega_2 \omega_3}
-{d_G\over 5760}{10(\omega_1^2 \omega_2^2+\omega_2^2 \omega_3^2+\omega_3^2 \omega_1^2)+7(\omega_1^4+\omega_2^4+\omega_3^4)\over \omega_1 \omega_2 \omega_3}
\end{align}
For all the above results of 6d (1,0) theories, we can see that our relation (\ref{eq925_conjecture}) is valid by comparing with the Cardy formula (\ref{eq766_estring}) and (\ref{eq568_results}).
\par 
Before ending this section, let us make a few comments on the relation between the Cardy formula and the supersymmetric Casimir energy. Under the presence of the Casimir energy, the index function \(Z\) can be divided into the Casimir energy part \(E_0\) and the spectral/state-counting part \(I\) as follows.
\begin{align}
Z=e^{- E_0} I
\end{align}
Casimir energy becomes the dominant contribution to \(\log Z\) when the chemical potentials are large. However, our Cardy formula determines the behavior of the spectral part \(\log I\) in the Cardy limit where the real parts of the chemical potentials are small. In 6d SCFTs, the Cardy formula and the Casimir energy have similar, but different expressions. Let us explain the difference between two quantities in the following paragraph.
\par 
As an example, let us consider the 6d (2,0) theory. Its supersymmetric Casimir energy is given by \cite{Bobev:2015kza}\footnote{The notations can be matched by \((2\beta \sigma_1)_\text{there}=(\Delta_R+\Delta_L)_\text{here}\), \((2\beta\sigma_2)_\text{there}=(\Delta_R-\Delta_L)_\text{here}\), and \((\beta\omega_I)_\text{there}=(\omega_I)_\text{here}\) }
\begin{align}
 &E_0=-{h_G^\vee d_G+r_G\over 384}{ (\Delta_R^2-\Delta_L^2)^2\over \omega_1\omega_2 \omega_3}
\nonumber \\
&-r_G \Bigr[{\Delta_R^2 \Delta_L^2\over 192 \omega_1 \omega_2 \omega_3}-{\Delta_R^2+\Delta_L^2\over 192} {\omega_1^2+\omega_2^2+\omega_3^2\over \omega_1 \omega_2 \omega_3}
+{\omega_1^4+\omega_2^4+\omega_3^4-2\omega_1^2 \omega_2^2-2\omega_2^2 \omega_3^2-2\omega_3^2 \omega_1^2\over 192\omega_1 \omega_2 \omega_3}\Bigr].
\label{eq1082_casimir}
\end{align}
In (\ref{eq1082_casimir}), the chemical potentials are defined in the conventional basis as (\ref{eq347_op}), which are constrained by \(\Delta_R-\omega_1-\omega_2-\omega_3=0\). On the other side, our Cardy formula is given by (\ref{eq410_cardy}) which is
\begin{align}
&\log I=-{h_G^\vee  d_G+r_G\over 384}{(\Delta_R^2-\Delta_L^2)^2\over \omega_1 \omega_2 \omega_3}
\nonumber \\
&-r_G\Bigr[{(\Delta_R^2 +4\pi^2)( \Delta_L^2+4\pi^2)\over 192\omega_1 \omega_2 \omega_3}-{\Delta_R^2+\Delta_L^2-8\pi^2\over 192}{\omega_1^2+\omega_2^2+\omega_3^2\over \omega_1 \omega_2 \omega_3} \Bigr]+\mathcal{O}(\log \omega).
\label{eq1091_cardy}
\end{align}
In (\ref{eq1091_cardy}), the chemical potentials are defined in the modified basis as (\ref{eq458_mod}), which are constrained by \(\Delta_R-\omega_1-\omega_2-\omega_3=2\pi i\).
\par 
Although the Casimir energy and the Cardy free energy are defined in the different basis, they share similar expressions. The similarity between the Casimir energy and the Cardy formula was also observed in \cite{Cabo-Bizet:2018ehj, Cabo-Bizet:2019osg, Kantor:2019lfo}. However, in 6d, the subleading corrections in the Cardy formula differ from the Casimir energy. More precisely, the homogeneous degree-one terms of the chemical potentials in the Cardy formula reproduce the Casimir energy up to \(\mathcal{O}(\omega^{-1})\). Also, the homogenous degree-one terms in the equivariant integral of the thermal anomaly polynomial (\ref{eq952_equiv}) reproduce the entire Casimir energy. In terms of the CS actions classified in (\ref{eq209_three}), the homogeneous terms come from \(W_\text{bulk}\) and \(W_\text{NCS}\) only. Those CS terms can be entirely determined from the ordinary anomaly polynomial \(P_8\), while the invariant CS term \(W_\text{ICS}\) needs the information of the thermal anomaly polynomial \(P_8^T\) to be determined. As a result, the Casimir energy coincides with the homogeneous terms in the Cardy formula up to \(\mathcal{O}(\omega^{-1})\).

\section{Asymptotic entropy}
\label{section5}
In this section, we consider the microcanonical ensemble of the 6d SCFTs and compute the asymptotic entropy in the Cardy limit. For 6d (2,0) SCFTs of \(A_N\) and \(D_N\) type, their large \(N\) free energies precisely account the entropy of BPS black holes in the dual \(AdS_7\). We further consider a general 6d (1,0) SCFTs and find a bound on the 't Hooft anomaly coefficients from the non-negativity condition of the entropy.
\subsection{Holographic SCFTs and \(AdS_7\) black holes}
 The large \(N\) limit of 6d (2,0) SCFTs are known to be dual to eleven-dimensional supergravity on \(AdS_7\times M^4\) where \(M^4\) should be \(S^4\) or its orbifolds \cite{Apruzzi:2013yva}. The internal manifold \(M^4\) is given by \(S^4\) for \(A_N\) type SCFT and \(S^4/\mathbb{Z}_2\) for \(D_N\) type SCFT. Here, we compute the degeneracies of BPS states in those holographic SCFTs and compare with the Bekenstein-Hawking entropy of BPS black holes in \(AdS_7\). The degeneracy \(\Omega\) of the BPS states in 6d (2,0) SCFTs can be obtained by the inverse Laplace transformation of the index \(I\). In the Cardy limit, it can be evaluated by the saddle point approximation which is given by the extremization of the following entropy function,
\begin{align}
S=\log I+\Delta_R Q_R+\Delta_L Q_L+\omega_1 J_1+\omega_2 J_2+\omega_3 J_3
\end{align}
where \(S=\log \Omega\) is the entropy of the BPS states. The extremization should be done over the chemical potentials constrained by \(\Delta_R-\omega_1-\omega_2-\omega_3=2\pi i\). 
\par 
Let us first discuss the large \(N\) limit of the 6d (2,0) \(A_{N}\) SCFT. In the Cardy limit, its free energy is given by,
\begin{align}
&\log I_{A_N}=-{(N+1)^3-1\over 384}{(\Delta_R^2-\Delta_L^2)^2\over \omega_1 \omega_2 \omega_3}
\nonumber \\
&-N\Bigr[{(\Delta_R^2 +4\pi^2)( \Delta_L^2+4\pi^2)\over 192\omega_1 \omega_2 \omega_3}-{\Delta_R^2+\Delta_L^2-8\pi^2\over 192}{\omega_1^2+\omega_2^2+\omega_3^2\over \omega_1 \omega_2 \omega_3} \Bigr]+\mathcal{O}(\log \omega).
\label{eq609_cardy}
\end{align}
In the remaining part of this section, we will assume that \(\mathcal{O}(\log \omega)\) correction is subleading than \(N^3\) in the large \(N\). Then, the large \(N\) free energy is
\begin{align}
\log I_{A_N}=-{N^3\over 384}{(\Delta_R^2-\Delta_L^2)^2\over \omega_1 \omega_2 \omega_3}.
\label{eq614_largeN}
\end{align}
In order to reproduce the black hole entropy from the above large \(N\) free energy, we should convert the field-theoretic quantity \(N\) to the gravitational quantity \(G_7\) which is the Newton constant of the dual \(AdS_7\). It can be computed from the dimensional reduction of the eleven-dimensional supergravity whose Netwon constant is given by \(G_{11}=16\pi^7 \ell_P^9\) where \(\ell_P\) is the Plank length. Since the 11d manifold is given by a direct product of \(AdS_7\) and \(S^4\), \(G_7\) can be obtained by
\begin{align}
G_7={G_{11}\over \text{vol}(S^4)}.
\end{align}
Let us denote \(\ell\) as a radius of \(AdS_7\). Then, the radius of \(S^4\) is given by \(\ell/2\). From the quantization of the 4-form flux on \(S^4\), \(\ell\) is related to \(N\) by \cite{Maldacena:1997re}
\begin{align}
N={1\over 8\pi}{\ell^3\over \ell_P^3}.
\end{align}
Therefore, the \(AdS_7\) Newton constant is given as follows,
\begin{align}
G_7=16\pi^7 \ell_P^9 \times \Bigr( {8\over 3}\pi^2 ({\ell\over 2})^4 \Bigr)^{-1}={3\pi^2 \ell^5\over 16N^3}.
\end{align}
In terms of \(G_7\), the free energy of \(A_{N}\) SCFT in the large \(N\) limit can be written as follows,
\begin{align}
\log I_{A_N}=-{\pi^2 \ell^5\over 2048 G_7}{(\Delta_R^2-\Delta_L^2)^2\over \omega_1 \omega_2 \omega_3}.
\label{eq519_AN}
\end{align}
The entropy can be obtained by extremizing the following entropy function,
\begin{align}
S=-{\pi^2 \ell^5\over 2048 G_7}{(\Delta_R^2-\Delta_L^2)^2\over \omega_1 \omega_2 \omega_3}+\Delta_R Q_R+\Delta_L Q_L+\omega_1 J_1+\omega_2 J_2+\omega_3 J_3 .
\label{eq663_ext}
\end{align}
The above extremization was first performed in \cite{Hosseini:2017mds}, and analytic expressions were derived in \cite{Choi:2018hmj}. The entropy \(S\) in (\ref{eq663_ext}) can be compared to the entropy of BPS black holes in asymptotic \(AdS_7\) whose solution was obtained in  \cite{Cvetic:2005zi}. The known BPS black hole solution satisfies a certain `charge relation' which comes from the fact that the BPS black holes exist on the intersection of BPS bound and extremal bound. After imposing the charge relation of the BPS black holes, we obtain the following expression for the entropy by extremizing (\ref{eq663_ext}),
 \begin{align}
 S&=2\pi \sqrt{3(Q_1^2 Q_2+Q_1 Q_2^2)-{3\pi^2 \ell^5\over 16G_7} (J_1 J_2+J_2 J_3+J_3 J_1)\over 3(Q_1+Q_2)-{3\pi^2 \ell^5\over 16G_7} }
 \nonumber \\
 &=2\pi \Bigr({\pi^2 \ell^5\over 16G_7}(J_1+J_2+J_3)+{Q_1^2+Q_2^2\over 2}+2Q_1 Q_2\Bigr)^{1\over 2}
 \nonumber \\
 &\times \Biggr(1-\sqrt{1-{{\pi^2 \ell^5\over 8G_7}J_1 J_2 J_3+Q_1^2 Q_2^2\over \Bigr({\pi^2 \ell^5\over 16G_7}(J_1+J_2+J_3)+{Q_1^2+Q_2^2\over 2}+2Q_1 Q_2\Bigr)^2 } } \Biggr) ^{1\over 2}
 \label{eq662_entropy}
 \end{align}
 where \(Q_1\equiv {Q_R+Q_L}\) and \(Q_2\equiv {Q_R-Q_L}\). The equivalence of two expressions in (\ref{eq662_entropy}) is the charge relation of BPS black holes. The obtained entropy (\ref{eq662_entropy}) is exactly the same with the entropy of an \(AdS_7\) black hole solution obtained in \cite{Cvetic:2005zi}. 
 \par 
Now, we consider the large \(N\) limit of \(D_N\) SCFT and its gravity dual. In the Cardy limit, the free energy of the \(D_N\) theory is
\begin{align}
&\log I_{D_N}=-{(2N-2)(2N^2-N)+N\over 384}{(\Delta_R^2-\Delta_L^2)^2\over \omega_1 \omega_2 \omega_3}
\nonumber \\
&-N\Bigr[{(\Delta_R^2 +4\pi^2)( \Delta_L^2+4\pi^2)\over 192\omega_1 \omega_2 \omega_3}-{\Delta_R^2+\Delta_L^2-8\pi^2\over 192}{\omega_1^2+\omega_2^2+\omega_3^2\over \omega_1 \omega_2 \omega_3} \Bigr]
+\mathcal{O}(\log \omega).
\end{align}
The large \(N\) free energy is
\begin{align}
\log I_{D_N}=-{N^3\over 96}{(\Delta_R^2-\Delta_L^2)^2\over \omega_1 \omega_2 \omega_3 }. 
\label{eq610_dn}
\end{align}
The corresponding gravity dual theory is eleven-dimensional supergravity on \(AdS_7\times S^4/\mathbb{Z}_2\). Due to \(\mathbb{Z}_2\) orbifold, \(\text{vol}(S^4/\mathbb{Z}_2)=\text{vol}(S^4)/2\). Also, the flux quantization on \(S^4/\mathbb{Z}_2\) is changed as
\begin{align}
N={1\over 16\pi}{\ell^3\over \ell_P^3}.
\end{align}
And the Newton constant on \(AdS_7\) is given by
\begin{align}
G_7={G_{11}\over \text{vol}(S^4/\mathbb{Z}_2)}={3\pi^2 \ell^5\over 64 N^3}.
\end{align}
Therefore the large \(N\) free energy of \(D_N\) theory can be written as
\begin{align}
\log I_{D_N}=-{\pi^2 \ell^5\over 2048G_7 }{(\Delta_R^2-\Delta_L^2)^2\over \omega_1 \omega_2 \omega_3 }.
\label{eq689_dn}
\end{align}
In \(AdS_7\) gravity, (\ref{eq689_dn}) has the same form with (\ref{eq519_AN}), thus accounting the same macroscopic entropy of the \(AdS_7\) black hole.
\par 
Before concluding this section, let us make some comments on the large \(N\) free energies and the Cardy limit. The obtained large \(N\) free energies (\ref{eq614_largeN}) and (\ref{eq610_dn}) are seemingly the leading order \(\mathcal{O}(\omega^{-3})\) in the Cardy limit, but actually they include the subleading terms up to \(\mathcal{O}(\omega^1)\) since \(\Delta_R=2\pi i +\omega_1+\omega_2+\omega_3\). Even considering \(\mathcal{O}(\log\omega)\) correction, (\ref{eq614_largeN}) and (\ref{eq610_dn}) are exact up \(\mathcal{O}(\omega^{-1})\). Therefore, our Cardy series expansion yields the correct entropy function of \(AdS_7\) black holes beyond the leading order.

\subsection{Bound on the anomaly coefficients}
In this section, we consider a general 6d (1,0) SCFT without flavor symmetry and analyze its entropy in the Cardy limit. Let us consider the following form of the modified superconformal index,
\begin{align}
I=\text{Tr}\Bigr[e^{-\Delta_R Q_R}\cdot e^{-\omega_1 J_1-\omega_2 J_2-\omega_3 J_3} \Bigr],\quad 
\Delta_R-\sum_{I=1}^3 \omega_I=2\pi i.
\end{align}
where all notations are the same with section \ref{section3}. The anomaly polynomial of the 6d SCFT with \(SU(2)_R\) R-symmetry and the tangent bundle can be written as follows,
\begin{align}
    P_8={1\over 4!}\Bigr(\mathfrak{a} \cdot c_2(R)^2+\mathfrak{b}\cdot c_2(R) p_1(T)+\mathfrak{c} \cdot p_1(T)^2+\mathfrak{d} \cdot p_2(T)  \Bigr)
    \label{eq992_anom}
\end{align}
where \(\mathfrak{a,b,c}\) and \(\mathfrak{d}\) are the anomaly coefficients of the theory. The Chern class and the Pontryagin classes are defined in (\ref{eq329_c2}) and (\ref{eq335_p12}). From the anomaly polynomial (\ref{eq992_anom}), we obtain the Cardy free energy (\ref{eq470_cardy}) as follows,
\begin{align}
\log I=&-\Bigr({\mathfrak{a}\over 384}\Delta_R^4 +{\mathfrak{b} \pi^2\over 24} \Delta_R^2+{2\mathfrak{c} \pi^4\over 3} \Bigr){1\over \omega_1 \omega_2 \omega_3}
+\Bigr({ \mathfrak{b}\over 96}\Delta_R^2 +{  (2\mathfrak{c}+\mathfrak{d}) \pi^2\over 6}\Bigr){\omega_1^2 +\omega_2^2+\omega_3^2\over \omega_1 \omega_2 \omega_3}+\mathcal{O}(\log \omega).
\end{align}
which was derived in section \ref{section2}. The asymptotic entropy of the SCFT can be obtained by extremizing the following entropy function with respect to the chemical potentials,
\begin{align}
S=\log I+\Delta_R Q_R+\sum_{I=1}^3\omega_I J_I
\end{align}
where chemical potentials are constrained by \(\Delta_R-\omega_1-\omega_2-\omega_3=2\pi i\). Once we impose this constraint, the entropy function becomes,
\begin{align}
&S=-{\pi^4 (\mathfrak{a}-4\mathfrak{b}+16\mathfrak{c})\over 24 \omega_1 \omega_2 \omega_3}
+\mathcal{O}(\omega^{-2}) +\sum_{I=1}^3 \omega_I (J_I+Q)+2\pi i Q
\end{align}
where we keep only the leading term in ther Cardy limit. Now, let us consider the equal chemical potential setting \(\omega\equiv \omega_{1,2,3}\) for the simplicity. Then, the entropy function is
\begin{align}
&S=-{\pi^4 (\mathfrak{a}-4\mathfrak{b}+16\mathfrak{c})\over 24 \omega^3 }
+\mathcal{O}(\omega^{-2})
+3\omega(J+Q)+2\pi i Q.
\label{eq1020_S}
\end{align}
where \(J\equiv J_{1,2,3}\). The extremization equation \({\partial S\over \partial \omega}=0\) determines the charge as the function of the chemical potentials as follows,
\begin{align}
 J+Q=-{\pi^4 (\mathfrak{a}-4\mathfrak{b}+16\mathfrak{c})\over 24\omega^4}+\mathcal{O}(\omega^{-3}).
\label{eq1025_ent}
\end{align}
Let us denote the real and the imaginary part of the chemical potential explicitly as \(\omega=\omega_R+i\omega_I\) where \(\omega_{I}\in\mathbb{R}\) and \(\omega_R>0\). Then, we can separate the real and the imaginary part of (\ref{eq1025_ent}) as follows,
\begin{align}
J+Q=-{\pi^4 (\mathfrak{a}-4\mathfrak{b}+16\mathfrak{c})\over 24|\omega|^8}\Bigr( (\omega_R^4-6\omega_R^2 \omega_I^2+\omega_I^6)+4i \omega_R \omega_I (\omega_I^2-\omega_R^2) \Bigr)+\mathcal{O}(\omega^{-3}).
\label{eq1030_JQ}
\end{align}
Since the charge should be real, we obtain \(\omega_I\) as a function of \(\omega_R\) as follows,
\begin{align}
\omega_I=\pm \omega_R +\mathcal{O}(\omega_R^2).
\label{eq1034_wi}
\end{align}
If we plug (\ref{eq1034_wi}) and (\ref{eq1030_JQ}) to (\ref{eq1020_S}), we obtain the real part of the entropy as follows,
\begin{align}
\text{Re}[S]={\pi^4(\mathfrak{a}-4\mathfrak{b}+16\mathfrak{c}) \over 24\omega_R^3}+\mathcal{O}(\omega_R^{-2}).
\label{eq1043_asympent}
\end{align}
As a result, the real part of the entropy (\ref{eq1043_asympent}) is non-negative as long as the anomaly coefficients satisfy the following bound,
\begin{align}
\mathfrak{a}-4\mathfrak{b}+16\mathfrak{c} \geq 0.
\label{eq1048_bound}
\end{align}
In table \ref{tab2}, we tabulated the anomaly coefficients of 6d free multiplets and various SCFTs. As can be seen, the entropy coefficient \(\mathfrak{a}-4\mathfrak{b}+16\mathfrak{c}\) is positive for all examples analyzed in this paper.

\begin{table}\hspace{-2.75cm} {\renewcommand{\arraystretch}{1.2}
\begin{tabular}{|c||c|c|c|c|c|}\hline
& \(\mathfrak{a}\) & \(\mathfrak{b}\) & \(\mathfrak{c}\) & \(\mathfrak{d}\) & \(\mathfrak{a}-4\mathfrak{b}+16\mathfrak{c}\) \\ \hline \hline
(1,0) tensor & \(1\) & \({1\over 2}\) & \({23\over 240}\) &\(-{29\over 60}\)  & \({8\over 15}\) \\ \hline
(1,0) hyper & \(0\) & \(0\) & \({7\over 240}\) & \(-{1\over 60}\) & \({7\over 15}\) \\ \hline
(2,0) tensor & \(1\) & \({1\over 2}\) & \({1\over 8}\) & \(-{1\over 2}\) & \(1\) \\ \hline
(2,0) ADE & \(h_G^\vee d_G+r_G\) & \({r_G\over 2}\) & \({r_G\over 8}\) & \(-{r_G\over 2}\) & \(h_G^\vee d_G +r_G\) \\ \hline
rank \(N\) E-string & \(4N^3-6N^2+3N\) & \(3N^2-{5\over 2}N\) & \({7\over 8}N\) &\(-{N\over 2}\) & \(4N^3-18N^2+27N\) \\ \hline
\(N\) M5's on \(\mathbb{C}^2/\Gamma_G\) &\makecell{\(|\Gamma|^2 N^3  \) \\ \(-2N((r_G+1)|\Gamma|-1)+d_G\)} & \(N(1-{r_G+1\over 2}|\Gamma|)+{d_G\over 2}\) & \({N\over 8}+{7d_G\over 240}\) & \(-{N\over 2}-{d_G\over 60}\) & \(|\Gamma|^2 N^3-{8d_G\over 15}\)\\ \hline
\end{tabular}}
\caption{Anomaly coefficients of free multiplets and various SCFTs in 6d.}
\label{tab2}
\end{table}

\acknowledgments

We thank Nikolay Bobev, Sunjin Choi, Joonho Kim, Jung-Wook Kim, Ki-Hong Lee, Jaewon Song, and Yang Zhou for the helpful discussion and comments. We especially appreciate Seok Kim for suggesting the project, giving inspiring comments, and carefully reading the manuscript. This work is supported by the National Research Foundation of Korea (NRF) Grant 2018R1A2B6004914, the Hyundai Motor Chung Mong-Koo Foundation.

\newpage

\appendix 

\section{A 4d \(\mathcal{N}=1\) Cardy formula }
In this section, we reproduce a 4d \(\mathcal{N}=1\) Cardy formula found in \cite{Kim:2019yrz, Cabo-Bizet:2019osg} by following our anomaly-based approach. We consider the following superconformal index of a 4d \(\mathcal{N}=1\) theory,
\begin{align}
I=\text{Tr}\Bigr[e^{-\Delta R}\cdot e^{-\omega_1 J_1-\omega_2 J_2} \Bigr],\quad \Bigr(2\Delta-\omega_1-\omega_2=2\pi i\Bigr)
\end{align}
where \(R\) is a charge of \(U(1)\) R-symmetry and \(J_{1,2}\) are angular momenta on \(S^3\). The background metric of \(S^3\times S^1\) is given by \cite{Choi:2018hmj}
\begin{align}
ds^2&=d\tau^2+r^2\Bigr[d\theta^2 +\sum_{i=1}^2 n_i^2 \theta \Bigr(d\phi_i-{i\omega_i\over \beta}d\tau \Bigr)^2 \Bigr],\quad (n_1,n_2)=(\cos\theta,\sin\theta)
\nonumber \\
&=e^{-2\Phi}(d\tau+\mathbf{a} )^2+r^2 \Bigr[d\theta^2+\sum_{i=1}^2 n_i^2 d\phi_i^2+{r^2\sum_i \omega_i n_i^2 d\phi_i \over \beta^2\Bigr(1-r^2 \sum_i {n_i^2 \omega_i^2\over \beta^2 } \Bigr)} \Bigr]
\end{align}
where \(r\) is a radius of \(S^3\) and \(\tau\sim\tau+\beta\). The dilaton \(\Phi\) and the graviphoton \(\mathbf{a}\) are
\begin{align}
e^{-2\Phi}=1-r^2\sum_{i=1}^2 {n_i^2 \omega_i^2\over \beta^2},\quad 
\mathbf{a}=-i{r^2 \sum_i \omega_i n_i^2 d\phi_i\over \beta\Bigr(1-r^2 \sum_i {n_i^2 \omega_i^2\over \beta^2} \Bigr)}
\end{align}
The background \(U(1)\) R-symmetry field is given by
\begin{align}
\mathbf{A}={\Delta\over \beta}d\tau
\end{align} 
The anomaly polynomial of 4d \(\mathcal{N}=1\) theories has the following form,
\begin{align}
P_6={1\over 6}k_{RRR}c_1(R)^3-{1\over 24}k_{R} c_1(R) p_1(T),\quad c_1(R)={i\over 2\pi}d\mathbf{A}
\end{align}
where \(c_1(R)\) is the first Chern-class of \(U(1)\) R-symmetry and \(p_1(T)\) is the first Pontryagin class of a tangent bundle. Two central charges \(a\) and \(c\) of 4d SCFTs are related to the 't Hooft anomaly coefficients as follows \cite{Anselmi:1997am},
\begin{align}
a={3\over 32}\Bigr(3k_{RRR}-k_{R}\Bigr),\quad c={1\over 32}\Bigr(9k_{RRR}-5k_{R} \Bigr)
\end{align}
Then the anomaly polynomial can be written in terms of the central charges as follows.
\begin{align}
P_6={8(5a-3c)\over 27}c_1(R)^3-{2(a-c)\over 3}c_1(R) p_1(T)
\label{eq1231_4d}
\end{align}
The thermal anomaly polynomial \(P_6^T\) can be obtained from the replacement rule (\ref{eq273_repla}) as follows,
\begin{align}
P_6^T={8(5a-3c)\over 27}c_1(R)^3-{2(a-c)\over 3}c_1(R) \Bigr(p_1(T)-{\mathbf{F}_T^2\over 4\pi^2}\Bigr)
\label{eq1250_4dtherm}
\end{align}
The CS action on \(S^3\) is
\begin{align}
W_\text{CS}=2\pi i \int_{\mathcal{N}^5} {dt+\mathbf{a}\over d\mathbf{a}} \Bigr[P^T_6-\hat{P}^T_6 \Bigr]_{\mathbf{A}^T=0}
\end{align}
\(\mathcal{N}^5=B^4\times S^1\) where \(B^4\) is a four-dimensional ball with \(\partial B^4=S^3\). After some calculation, we obtain the following CS action,
\begin{align}
W_\text{CS}&=\Bigr({2(5a-3c)\over 27\pi^2}\Delta^3 +{2(a-c)\over 3}\Delta\Bigr){1\over \beta^2}\int_{S^3}\mathbf{a}d\mathbf{a}
-{a-c\over 3(2\pi)^2}\Delta \int_{S^3} \text{tr}\Bigr[\hat{\mathbf{\Gamma}}d\hat{\mathbf{\Gamma}}+{2\over 3}\hat{\mathbf{\Gamma}}^3 \Bigr]
\end{align}
There are only gauge and mixed CS terms on \(S^3\). After integrating over \(S^3\), we obtain the following results,
\begin{align}
&{1\over \beta^2}\int_{S^3} \mathbf{a}d\mathbf{a}=-{(2\pi)^2 \omega_1 \omega_2\over \Bigr({\beta^2\over r^2}-\omega_1^2 \Bigr)   \Bigr({\beta^2\over r^2}-\omega_2^2 \Bigr) }=-{4\pi^2 \over \omega_1 \omega_2}\cdot \Bigr[1+\mathcal{O}\Bigr({\beta^2\over r^2 \omega^2} \Bigr) \Bigr]
\nonumber \\
&\int_{S^3}\text{tr}\Bigr[\hat{\mathbf{\Gamma}}d\hat{\mathbf{\Gamma}}+{2\over 3}\hat{\mathbf{\Gamma}}^3 \Bigr]={\beta^2\over r^2}{16\pi^2 \omega_1 \omega_2\over \Bigr({\beta^2\over r^2}-\omega_1^2\Bigr)\Bigr({\beta^2\over r^2}-\omega_2^2\Bigr) }=\mathcal{O}\Bigr({\beta^2\over r^2 \omega^2} \Bigr).
\end{align}
Therefore in the scaling limit (\ref{eq296_scaling}), the Cardy free energy becomes
\begin{align}
\log I&=-\lim_{\beta\to 0}W_\text{CS}+\mathcal{O}(\log \omega)
={8\Delta^3\over 27 \omega_1 \omega_2}(5a-3c)+{8\pi^2 \Delta\over 3\omega_1 \omega_2} (a-c)+\mathcal{O}(\log \omega)
\label{eq1271_4dcardy}
\end{align}
It is exactly the same with the Cardy formula given in \cite{Kim:2019yrz, Cabo-Bizet:2019osg}. 
\par 
The Cardy formula (\ref{eq1271_4dcardy}) can also be obtained from the equivariant integral of the thermal anomaly polynomial (\ref{eq1250_4dtherm}) as follows,
\begin{align}
\log I =-\int P_6^T+\mathcal{O}(\log \omega)
\label{eq1423_conje}
\end{align}
Now, we use the Duistermaat-Heckman formula (\ref{eq1013_DH}) to evaluate the equivariant integral. The 0-form components of the equivariant classes at the fixed point \(p_0\) are given as follows,
\begin{align}
c_1(R)_{p_0} = -\Delta,\quad 
p_1(T)|_{p_0}= \omega_1^2+\omega_2^2,\quad 
e(TM)|_{p_0}=\omega_1 \omega_2,\quad 
{\mathbf{F}_T^2\over 4\pi^2}\Bigr|_{p_0} = -(2\pi i)^2
\end{align}
Then, we obtain
\begin{align}
-\int P_6^T={8\Delta^3\over 27\omega_1 \omega_2}(5a-3c)+{8\pi^2 \Delta\over 3\omega_1 \omega_2 }(a-c) -{2\Delta (\omega_1^2+\omega_2^2)\over 3\omega_1 \omega_2}(a-c).
\label{eq1283_equiv}
\end{align}
As a result, the Cardy formula (\ref{eq1271_4dcardy}) can be obtained from (\ref{eq1283_equiv}) by our relation (\ref{eq1423_conje}). In this case, \(\mathcal{O}(\omega^0)\) term in (\ref{eq1283_equiv}) can be ignored since it is subleading than \(\mathcal{O}(\log \omega)\).

\section{Modified index of 6d supermultiplets}
\label{appendix_B}
In this section, we compute the modified superconformal indices of free (1,0) tensor and hypermultiplets defined as follows,
\begin{align}
    I=\text{Tr}\Bigr[e^{-2\pi iQ_R}\cdot e^{-\omega_1 (J_1+Q_R)-\omega_2 (J_2+Q_R)-\omega_3 (J_3+Q_R)} \Bigr].
\end{align}
Note that \(\Delta_R\) in (\ref{eq725_modind}) is absorbed by the chemical potential relation \(\Delta_R-\omega_1-\omega_2-\omega_3=2\pi i\). More detailed explanation is given in section \ref{section3}.
\par 
6d (1,0) hypermultiplet has component fields \((\phi,\psi^+)\) where \(\phi\) is a complex scalar in \((\mathbf{2},\mathbf{1})\) representation of \(SU(2)_R\times SO(6)\), and \(\psi^+\) is a fermion in \((\mathbf{1},\mathbf{ 4})\). Tensor multiplet has component fields \((a,\xi^+,B_{\mu\nu}^+)\) where \(a\) is a real scalar in \((\mathbf{1},\mathbf{1})\), \(\xi^+\) is a fermion in \((\mathbf{2},\mathbf{ 4})\), and \(B_{\mu\nu}^+\) is a self-dual tensor in \((\mathbf{1},\mathbf{10 })\). The charges of the BPS fields that saturate the BPS bound \(E\geq J_1+J_2+J_3+4Q_R\) are listed in the table \ref{table1373}.

\begin{table}
\centering
\begin{tabular}{|c|c|c|c|c|c|c|c|c|}
\hline
multiplet & letter & type & \(E\) & \(Q_R\) & \(J_1\) & \(J_2\) & \(J_3\)  & letter index \\ \hline \hline
half-hyper & \(\phi\) & B & \(2\) & \(1/2\) & 0 & 0 & 0 & \((-1)q_1 q_2 q_3\) \\ \hline  
tensor & \(B_{\mu\nu}^+\) & B & 3 & 0 & 1 & 1 & 1 & \( (q_1 q_2 q_3)^2 \) \\ \hline
tensor & \(\xi^+\) & F & \(5/2\) & \(1/2\) & \(1/2\) & \(1/2\) & \(-1/2\) & \((-1)(q_1  q_2)^2 \) \\ \hline
tensor & \(\xi^+\) & F & \(5/2\) & \(1/2\) & \(1/2\) & \(-1/2\) & \(1/2\) & \((-1)(q_1 q_3)^2 \) \\ \hline
tensor & \(\xi^+\) & F & \(5/2\) & \(1/2\) & \(-1/2\) & \(1/2\) & \(1/2\) & \((-1)(q_2 q_3)^2 \) \\ \hline 
derivative & \(\partial_1\) & B & \(1\) & \(0\) & \(1\) & \(0\) & \(0\) & \(q_1^2\) \\ \hline 
derivative & \(\partial_2\) & B & \(1\) & \(0\) & \(0\) & \(1\) & \(0\) & \(q_2^2\) \\ \hline 
derivative & \(\partial_3\) & B & \(1\) & \(0\) & \(0\) & \(0\) & \(1\) & \(q_3^2\) \\ \hline 
\end{tabular}
\caption{BPS letters in 6d multiplets. Type denotes a boson/fermion statistics, and \(f\) denotes a single particle index of the BPS letter. The fugacities are defined as \(q_I=e^{-\omega_I/2}\). Note that \((-1)\) in the single particle index should be also treated as a fugacity.}
\label{table1373}
\end{table}

\par 
The superconformal index can be obtained from the plethystic exponential of the BPS letters as follows,
\begin{align}
I=\exp\sum_{n=1}^\infty {1\over n}\Bigr[f_B( \bullet^n)-(-1)^{n} f_F(\bullet^n) \Bigr]
\end{align}
where \(f_B\) is a single particle index of bosonic fields and \(f_B\) is for fermionic fields. Here, \(\bullet\) collectively denotes the fugacity variables. Then, the modified superconformal indices of (1,0) supermultiplets are given as follows,
\begin{align}
I_\text{hyp}&=\exp\sum_{n=1}^\infty{2\over n}{(-q_1 q_2 q_3)^n\over (1-q_1^{2n})(1-q_2^{2n})(1-q_3^{2n})}
\nonumber \\
I_\text{ten}&=\exp\sum_{n=1}^\infty{1\over n}{ (q_1 q_2 q_3)^{2n}-(q_1 q_2)^{2n}-(q_2 q_3)^{2n}-(q_3 q_1)^{2n} \over (1-q_1^{2n})(1-q_2^{2n})(1-q_3^{2n})}
\end{align}
where fugacities are defined as \(q_I=e^{-\omega_I/2}\). Note that by shifting \(\omega_3\to \omega_3-2\pi i\) which is equivalent to shifting \(q_3\to -q_3\), we obtain the original superconformal indices of tensor/hypermultiplets computed in \cite{Benvenuti:2016dcs}.

% The bibliography will probably be heavily edited during typesetting.
% We'll parse it and, using the arxiv number or the journal data, will
% query inspire, trying to verify the data (this will probalby spot
% eventual typos) and retrive the document DOI and eventual errata.
% We however suggest to always provide author, title and journal data:
% in short all the informations that clearly identify a document.

\newpage

\bibliographystyle{JHEP}
\bibliography{main}

\providecommand{\href}[2]{#2}\begingroup\raggedright\begin{thebibliography}{10}

\bibitem{Kinney:2005ej}
J.~Kinney, J.~M. Maldacena, S.~Minwalla and S.~Raju, \emph{{An Index for 4
  dimensional super conformal theories}},
  \href{https://doi.org/10.1007/s00220-007-0258-7}{\emph{Commun. Math. Phys.}
  {\bfseries 275} (2007) 209}
  [\href{https://arxiv.org/abs/hep-th/0510251}{{\ttfamily hep-th/0510251}}].

\bibitem{Romelsberger:2005eg}
C.~Romelsberger, \emph{{Counting chiral primaries in N = 1, d=4 superconformal
  field theories}},
  \href{https://doi.org/10.1016/j.nuclphysb.2006.03.037}{\emph{Nucl. Phys.}
  {\bfseries B747} (2006) 329}
  [\href{https://arxiv.org/abs/hep-th/0510060}{{\ttfamily hep-th/0510060}}].

\bibitem{Bhattacharya:2008zy}
J.~Bhattacharya, S.~Bhattacharyya, S.~Minwalla and S.~Raju, \emph{{Indices for
  Superconformal Field Theories in 3,5 and 6 Dimensions}},
  \href{https://doi.org/10.1088/1126-6708/2008/02/064}{\emph{JHEP} {\bfseries
  02} (2008) 064} [\href{https://arxiv.org/abs/0801.1435}{{\ttfamily
  0801.1435}}].

\bibitem{Kim:2012ava}
H.-C. Kim and S.~Kim, \emph{{M5-branes from gauge theories on the 5-sphere}},
  \href{https://doi.org/10.1007/JHEP05(2013)144}{\emph{JHEP} {\bfseries 05}
  (2013) 144} [\href{https://arxiv.org/abs/1206.6339}{{\ttfamily 1206.6339}}].

\bibitem{Kallen:2012zn}
J.~Kallen, J.~A. Minahan, A.~Nedelin and M.~Zabzine, \emph{{$N^3$-behavior from
  5D Yang-Mills theory}},
  \href{https://doi.org/10.1007/JHEP10(2012)184}{\emph{JHEP} {\bfseries 10}
  (2012) 184} [\href{https://arxiv.org/abs/1207.3763}{{\ttfamily 1207.3763}}].

\bibitem{Kim:2012tr}
H.-C. Kim and K.~Lee, \emph{{Supersymmetric M5 Brane Theories on R x CP2}},
  \href{https://doi.org/10.1007/JHEP07(2013)072}{\emph{JHEP} {\bfseries 07}
  (2013) 072} [\href{https://arxiv.org/abs/1210.0853}{{\ttfamily 1210.0853}}].

\bibitem{Lockhart:2012vp}
G.~Lockhart and C.~Vafa, \emph{{Superconformal Partition Functions and
  Non-perturbative Topological Strings}},
  \href{https://doi.org/10.1007/JHEP10(2018)051}{\emph{JHEP} {\bfseries 10}
  (2018) 051} [\href{https://arxiv.org/abs/1210.5909}{{\ttfamily 1210.5909}}].

\bibitem{Kim:2012qf}
H.-C. Kim, J.~Kim and S.~Kim, \emph{{Instantons on the 5-sphere and
  M5-branes}},  \href{https://arxiv.org/abs/1211.0144}{{\ttfamily 1211.0144}}.

\bibitem{Minahan:2013jwa}
J.~A. Minahan, A.~Nedelin and M.~Zabzine, \emph{{5D super Yang-Mills theory and
  the correspondence to AdS$_7$/CFT$_6$}},
  \href{https://doi.org/10.1088/1751-8113/46/35/355401}{\emph{J. Phys.}
  {\bfseries A46} (2013) 355401}
  [\href{https://arxiv.org/abs/1304.1016}{{\ttfamily 1304.1016}}].

\bibitem{Kim:2013nva}
H.-C. Kim, S.~Kim, S.-S. Kim and K.~Lee, \emph{{The general M5-brane
  superconformal index}},  \href{https://arxiv.org/abs/1307.7660}{{\ttfamily
  1307.7660}}.

\bibitem{Kim:2016usy}
S.~Kim and K.~Lee, \emph{{Indices for 6 dimensional superconformal field
  theories}}, \href{https://doi.org/10.1088/1751-8121/aa5cbf}{\emph{J. Phys.}
  {\bfseries A50} (2017) 443017}
  [\href{https://arxiv.org/abs/1608.02969}{{\ttfamily 1608.02969}}].

\bibitem{Klebanov:1996un}
I.~R. Klebanov and A.~A. Tseytlin, \emph{{Entropy of near extremal black
  p-branes}}, \href{https://doi.org/10.1016/0550-3213(96)00295-7}{\emph{Nucl.
  Phys.} {\bfseries B475} (1996) 164}
  [\href{https://arxiv.org/abs/hep-th/9604089}{{\ttfamily hep-th/9604089}}].

\bibitem{Freed:1998tg}
D.~Freed, J.~A. Harvey, R.~Minasian and G.~W. Moore, \emph{{Gravitational
  anomaly cancellation for M theory five-branes}},
  \href{https://doi.org/10.4310/ATMP.1998.v2.n3.a8}{\emph{Adv. Theor. Math.
  Phys.} {\bfseries 2} (1998) 601}
  [\href{https://arxiv.org/abs/hep-th/9803205}{{\ttfamily hep-th/9803205}}].

\bibitem{Ohmori:2014kda}
K.~Ohmori, H.~Shimizu, Y.~Tachikawa and K.~Yonekura, \emph{{Anomaly polynomial
  of general 6d SCFTs}}, \href{https://doi.org/10.1093/ptep/ptu140}{\emph{PTEP}
  {\bfseries 2014} (2014) 103B07}
  [\href{https://arxiv.org/abs/1408.5572}{{\ttfamily 1408.5572}}].

\bibitem{Intriligator:2014eaa}
K.~Intriligator, \emph{{6d, $ \mathcal{N}=\left(1,\;0\right) $ Coulomb branch
  anomaly matching}},
  \href{https://doi.org/10.1007/JHEP10(2014)162}{\emph{JHEP} {\bfseries 10}
  (2014) 162} [\href{https://arxiv.org/abs/1408.6745}{{\ttfamily 1408.6745}}].

\bibitem{Bobev:2015kza}
N.~Bobev, M.~Bullimore and H.-C. Kim, \emph{{Supersymmetric Casimir Energy and
  the Anomaly Polynomial}},
  \href{https://doi.org/10.1007/JHEP09(2015)142}{\emph{JHEP} {\bfseries 09}
  (2015) 142} [\href{https://arxiv.org/abs/1507.08553}{{\ttfamily
  1507.08553}}].

\bibitem{Kim:2011mv}
H.-C. Kim, S.~Kim, E.~Koh, K.~Lee and S.~Lee, \emph{{On instantons as
  Kaluza-Klein modes of M5-branes}},
  \href{https://doi.org/10.1007/JHEP12(2011)031}{\emph{JHEP} {\bfseries 12}
  (2011) 031} [\href{https://arxiv.org/abs/1110.2175}{{\ttfamily 1110.2175}}].

\bibitem{Kim:2017zyo}
S.~Kim and J.~Nahmgoong, \emph{{Asymptotic M5-brane entropy from S-duality}},
  \href{https://doi.org/10.1007/JHEP12(2017)120}{\emph{JHEP} {\bfseries 12}
  (2017) 120} [\href{https://arxiv.org/abs/1702.04058}{{\ttfamily
  1702.04058}}].

\bibitem{Hosseini:2017mds}
S.~M. Hosseini, K.~Hristov and A.~Zaffaroni, \emph{{An extremization principle
  for the entropy of rotating BPS black holes in AdS$_{5}$}},
  \href{https://doi.org/10.1007/JHEP07(2017)106}{\emph{JHEP} {\bfseries 07}
  (2017) 106} [\href{https://arxiv.org/abs/1705.05383}{{\ttfamily
  1705.05383}}].

\bibitem{Hosseini:2018dob}
S.~M. Hosseini, K.~Hristov and A.~Zaffaroni, \emph{{A note on the entropy of
  rotating BPS AdS$_7\times S^4$ black holes}},
  \href{https://doi.org/10.1007/JHEP05(2018)121}{\emph{JHEP} {\bfseries 05}
  (2018) 121} [\href{https://arxiv.org/abs/1803.07568}{{\ttfamily
  1803.07568}}].

\bibitem{Choi:2018fdc}
S.~Choi, C.~Hwang, S.~Kim and J.~Nahmgoong, \emph{{Entropy functions of BPS
  black holes in AdS$_4$ and AdS$_6$}},
  \href{https://arxiv.org/abs/1811.02158}{{\ttfamily 1811.02158}}.

\bibitem{Choi:2019zpz}
S.~Choi, C.~Hwang and S.~Kim, \emph{{Quantum vortices, M2-branes and black
  holes}},  \href{https://arxiv.org/abs/1908.02470}{{\ttfamily 1908.02470}}.

\bibitem{Nian:2019pxj}
J.~Nian and L.~A. Pando~Zayas, \emph{{Microscopic Entropy of Rotating
  Electrically Charged AdS$_4$ Black Holes from Field Theory Localization}},
  \href{https://arxiv.org/abs/1909.07943}{{\ttfamily 1909.07943}}.

\bibitem{Choi:2019dfu}
S.~Choi and C.~Hwang, \emph{{Universal 3d Cardy Block and Black Hole Entropy}},
   \href{https://arxiv.org/abs/1911.01448}{{\ttfamily 1911.01448}}.

\bibitem{Cabo-Bizet:2018ehj}
A.~Cabo-Bizet, D.~Cassani, D.~Martelli and S.~Murthy, \emph{{Microscopic origin
  of the Bekenstein-Hawking entropy of supersymmetric AdS$_{\bf 5}$ black
  holes}},  \href{https://arxiv.org/abs/1810.11442}{{\ttfamily 1810.11442}}.

\bibitem{Choi:2018hmj}
S.~Choi, J.~Kim, S.~Kim and J.~Nahmgoong, \emph{{Large AdS black holes from
  QFT}},  \href{https://arxiv.org/abs/1810.12067}{{\ttfamily 1810.12067}}.

\bibitem{Benini:2018ywd}
F.~Benini and P.~Milan, \emph{{Black holes in 4d $\mathcal{N}=4$
  Super-Yang-Mills}},  \href{https://arxiv.org/abs/1812.09613}{{\ttfamily
  1812.09613}}.

\bibitem{Choi:2019miv}
S.~Choi and S.~Kim, \emph{{Large AdS$_6$ black holes from CFT$_5$}},
  \href{https://arxiv.org/abs/1904.01164}{{\ttfamily 1904.01164}}.

\bibitem{Choi:2018vbz}
S.~Choi, J.~Kim, S.~Kim and J.~Nahmgoong, \emph{{Comments on deconfinement in
  AdS/CFT}},  \href{https://arxiv.org/abs/1811.08646}{{\ttfamily 1811.08646}}.

\bibitem{Cassani:2019mms}
D.~Cassani and L.~Papini, \emph{{The BPS limit of rotating AdS black hole
  thermodynamics}},  \href{https://arxiv.org/abs/1906.10148}{{\ttfamily
  1906.10148}}.

\bibitem{Kantor:2019lfo}
G.~Kántor, C.~Papageorgakis and P.~Richmond, \emph{{AdS$_7$ Black-Hole Entropy
  and 5D $\mathcal{N}=2$ Yang-Mills}},
  \href{https://arxiv.org/abs/1907.02923}{{\ttfamily 1907.02923}}.

\bibitem{Cardy:1986ie}
J.~L. Cardy, \emph{{Operator Content of Two-Dimensional Conformally Invariant
  Theories}}, \href{https://doi.org/10.1016/0550-3213(86)90552-3}{\emph{Nucl.
  Phys.} {\bfseries B270} (1986) 186}.

\bibitem{Strominger:1996sh}
A.~Strominger and C.~Vafa, \emph{{Microscopic origin of the Bekenstein-Hawking
  entropy}}, \href{https://doi.org/10.1016/0370-2693(96)00345-0}{\emph{Phys.
  Lett.} {\bfseries B379} (1996) 99}
  [\href{https://arxiv.org/abs/hep-th/9601029}{{\ttfamily hep-th/9601029}}].

\bibitem{DiPietro:2014bca}
L.~Di~Pietro and Z.~Komargodski, \emph{{Cardy formulae for SUSY theories in $d
  =$ 4 and $d =$ 6}},
  \href{https://doi.org/10.1007/JHEP12(2014)031}{\emph{JHEP} {\bfseries 12}
  (2014) 031} [\href{https://arxiv.org/abs/1407.6061}{{\ttfamily 1407.6061}}].

\bibitem{Honda:2019cio}
M.~Honda, \emph{{Quantum Black Hole Entropy from 4d Supersymmetric Cardy
  formula}},  \href{https://arxiv.org/abs/1901.08091}{{\ttfamily 1901.08091}}.

\bibitem{ArabiArdehali:2019tdm}
A.~Arabi~Ardehali, \emph{{Cardy-like asymptotics of the 4d $\mathcal{N}=4$
  index and AdS$_5$ blackholes}},
  \href{https://arxiv.org/abs/1902.06619}{{\ttfamily 1902.06619}}.

\bibitem{Kim:2019yrz}
J.~Kim, S.~Kim and J.~Song, \emph{{A 4d N=1 Cardy Formula}},
  \href{https://arxiv.org/abs/1904.03455}{{\ttfamily 1904.03455}}.

\bibitem{Cabo-Bizet:2019osg}
A.~Cabo-Bizet, D.~Cassani, D.~Martelli and S.~Murthy, \emph{{The asymptotic
  growth of states of the 4d N=1 superconformal index}},
  \href{https://arxiv.org/abs/1904.05865}{{\ttfamily 1904.05865}}.

\bibitem{Chang:2019uag}
C.-M. Chang, M.~Fluder, Y.-H. Lin and Y.~Wang, \emph{{Proving the 6d Cardy
  Formula and Matching Global Gravitational Anomalies}},
  \href{https://arxiv.org/abs/1910.10151}{{\ttfamily 1910.10151}}.

\bibitem{Jensen:2012kj}
K.~Jensen, R.~Loganayagam and A.~Yarom, \emph{{Thermodynamics, gravitational
  anomalies and cones}},
  \href{https://doi.org/10.1007/JHEP02(2013)088}{\emph{JHEP} {\bfseries 02}
  (2013) 088} [\href{https://arxiv.org/abs/1207.5824}{{\ttfamily 1207.5824}}].

\bibitem{Jensen:2013kka}
K.~Jensen, R.~Loganayagam and A.~Yarom, \emph{{Anomaly inflow and thermal
  equilibrium}}, \href{https://doi.org/10.1007/JHEP05(2014)134}{\emph{JHEP}
  {\bfseries 05} (2014) 134} [\href{https://arxiv.org/abs/1310.7024}{{\ttfamily
  1310.7024}}].

\bibitem{Jensen:2013rga}
K.~Jensen, R.~Loganayagam and A.~Yarom, \emph{{Chern-Simons terms from thermal
  circles and anomalies}},
  \href{https://doi.org/10.1007/JHEP05(2014)110}{\emph{JHEP} {\bfseries 05}
  (2014) 110} [\href{https://arxiv.org/abs/1311.2935}{{\ttfamily 1311.2935}}].

\bibitem{Yankielowicz:2017xkf}
S.~Yankielowicz and Y.~Zhou, \emph{{Supersymmetric Rényi entropy and Anomalies
  in 6d (1,0) SCFTs}},
  \href{https://doi.org/10.1007/JHEP04(2017)128}{\emph{JHEP} {\bfseries 04}
  (2017) 128} [\href{https://arxiv.org/abs/1702.03518}{{\ttfamily
  1702.03518}}].

\bibitem{tHooft:1977nqb}
G.~'t~Hooft, \emph{{On the Phase Transition Towards Permanent Quark
  Confinement}},
  \href{https://doi.org/10.1016/0550-3213(78)90153-0}{\emph{Nucl. Phys.}
  {\bfseries B138} (1978) 1}.

\bibitem{Wess:1971yu}
J.~Wess and B.~Zumino, \emph{{Consequences of anomalous Ward identities}},
  \href{https://doi.org/10.1016/0370-2693(71)90582-X}{\emph{Phys. Lett.}
  {\bfseries 37B} (1971) 95}.

\bibitem{Bilal:2008qx}
A.~Bilal, \emph{{Lectures on Anomalies}},
  \href{https://arxiv.org/abs/0802.0634}{{\ttfamily 0802.0634}}.

\bibitem{Bardeen:1984pm}
W.~A. Bardeen and B.~Zumino, \emph{{Consistent and Covariant Anomalies in Gauge
  and Gravitational Theories}},
  \href{https://doi.org/10.1016/0550-3213(84)90322-5}{\emph{Nucl. Phys.}
  {\bfseries B244} (1984) 421}.

\bibitem{Nahm:1977tg}
W.~Nahm, \emph{{Supersymmetries and their Representations}},
  \href{https://doi.org/10.1016/0550-3213(78)90218-3}{\emph{Nucl. Phys.}
  {\bfseries B135} (1978) 149}.

\bibitem{Witten:1995zh}
E.~Witten, \emph{{Some comments on string dynamics}},  in \emph{{Future
  perspectives in string theory. Proceedings, Conference, Strings'95, Los
  Angeles, USA, March 13-18, 1995}}, pp.~501--523, 1995,
  \href{https://arxiv.org/abs/hep-th/9507121}{{\ttfamily hep-th/9507121}}.

\bibitem{Seiberg:1996vs}
N.~Seiberg and E.~Witten, \emph{{Comments on string dynamics in
  six-dimensions}},
  \href{https://doi.org/10.1016/0550-3213(96)00189-7}{\emph{Nucl. Phys.}
  {\bfseries B471} (1996) 121}
  [\href{https://arxiv.org/abs/hep-th/9603003}{{\ttfamily hep-th/9603003}}].

\bibitem{Strominger:1995ac}
A.~Strominger, \emph{{Open p-branes}},
  \href{https://doi.org/10.1016/0370-2693(96)00712-5}{\emph{Phys. Lett.}
  {\bfseries B383} (1996) 44}
  [\href{https://arxiv.org/abs/hep-th/9512059}{{\ttfamily hep-th/9512059}}].

\bibitem{Fredenhagen:2004cj}
S.~Fredenhagen and V.~Schomerus, \emph{{Boundary Liouville theory at c = 1}},
  \href{https://doi.org/10.1088/1126-6708/2005/05/025}{\emph{JHEP} {\bfseries
  05} (2005) 025} [\href{https://arxiv.org/abs/hep-th/0409256}{{\ttfamily
  hep-th/0409256}}].

\bibitem{Heckman:2015bfa}
J.~J. Heckman, D.~R. Morrison, T.~Rudelius and C.~Vafa, \emph{{Atomic
  Classification of 6D SCFTs}},
  \href{https://doi.org/10.1002/prop.201500024}{\emph{Fortsch. Phys.}
  {\bfseries 63} (2015) 468}
  [\href{https://arxiv.org/abs/1502.05405}{{\ttfamily 1502.05405}}].

\bibitem{Horava:1995qa}
P.~Horava and E.~Witten, \emph{{Heterotic and type I string dynamics from
  eleven-dimensions}},
  \href{https://doi.org/10.1016/0550-3213(95)00621-4}{\emph{Nucl. Phys.}
  {\bfseries B460} (1996) 506}
  [\href{https://arxiv.org/abs/hep-th/9510209}{{\ttfamily hep-th/9510209}}].

\bibitem{Klemm:1996hh}
A.~Klemm, P.~Mayr and C.~Vafa, \emph{{BPS states of exceptional noncritical
  strings}}, \href{https://doi.org/10.1016/S0920-5632(97)00422-2}{\emph{Nucl.
  Phys. Proc. Suppl.} {\bfseries 58} (1997) 177}
  [\href{https://arxiv.org/abs/hep-th/9607139}{{\ttfamily hep-th/9607139}}].

\bibitem{DelZotto:2014hpa}
M.~Del~Zotto, J.~J. Heckman, A.~Tomasiello and C.~Vafa, \emph{{6d Conformal
  Matter}}, \href{https://doi.org/10.1007/JHEP02(2015)054}{\emph{JHEP}
  {\bfseries 02} (2015) 054} [\href{https://arxiv.org/abs/1407.6359}{{\ttfamily
  1407.6359}}].

\bibitem{Benvenuti:2016dcs}
S.~Benvenuti, G.~Bonelli, M.~Ronzani and A.~Tanzini, \emph{{Symmetry
  enhancements via 5d instantons, $ q\mathcal{W} $ -algebrae and (1, 0)
  superconformal index}},
  \href{https://doi.org/10.1007/JHEP09(2016)053}{\emph{JHEP} {\bfseries 09}
  (2016) 053} [\href{https://arxiv.org/abs/1606.03036}{{\ttfamily
  1606.03036}}].

\bibitem{Duistermaat:1982vw}
J.~J. Duistermaat and G.~J. Heckman, \emph{{On the Variation in the cohomology
  of the symplectic form of the reduced phase space}},
  \href{https://doi.org/10.1007/BF01399506}{\emph{Invent. Math.} {\bfseries 69}
  (1982) 259}.

\bibitem{Apruzzi:2013yva}
F.~Apruzzi, M.~Fazzi, D.~Rosa and A.~Tomasiello, \emph{{All $AdS_7$ solutions
  of type II supergravity}},
  \href{https://doi.org/10.1007/JHEP04(2014)064}{\emph{JHEP} {\bfseries 04}
  (2014) 064} [\href{https://arxiv.org/abs/1309.2949}{{\ttfamily 1309.2949}}].

\bibitem{Maldacena:1997re}
J.~M. Maldacena, \emph{{The Large N limit of superconformal field theories and
  supergravity}}, \href{https://doi.org/10.1023/A:1026654312961,
  10.4310/ATMP.1998.v2.n2.a1}{\emph{Int. J. Theor. Phys.} {\bfseries 38} (1999)
  1113} [\href{https://arxiv.org/abs/hep-th/9711200}{{\ttfamily
  hep-th/9711200}}].

\bibitem{Cvetic:2005zi}
M.~Cvetic, G.~W. Gibbons, H.~Lu and C.~N. Pope, \emph{{Rotating black holes in
  gauged supergravities: Thermodynamics, supersymmetric limits, topological
  solitons and time machines}},
  \href{https://arxiv.org/abs/hep-th/0504080}{{\ttfamily hep-th/0504080}}.

\bibitem{Anselmi:1997am}
D.~Anselmi, D.~Z. Freedman, M.~T. Grisaru and A.~A. Johansen,
  \emph{{Nonperturbative formulas for central functions of supersymmetric gauge
  theories}}, \href{https://doi.org/10.1016/S0550-3213(98)00278-8}{\emph{Nucl.
  Phys.} {\bfseries B526} (1998) 543}
  [\href{https://arxiv.org/abs/hep-th/9708042}{{\ttfamily hep-th/9708042}}].

\end{thebibliography}\endgroup

\end{document}